\documentclass[aoas]{imsart}

\RequirePackage{amsthm,amsmath,amsfonts,amssymb}
\RequirePackage[authoryear]{natbib}
\RequirePackage[colorlinks,citecolor=blue,urlcolor=blue]{hyperref}
\RequirePackage{graphicx}

\startlocaldefs
\theoremstyle{plain}

\theoremstyle{remark}

\endlocaldefs

\usepackage{epsfig}
\usepackage{graphicx,psfrag}
\usepackage{amssymb, amsmath}
\usepackage{amsthm,bm}
\usepackage{verbatim,multirow,rotating,mathtools}
\usepackage{float}
\usepackage{url}
\usepackage{color,colortbl}
\usepackage{caption}
\usepackage{subcaption}
\usepackage{multirow}
\usepackage{lscape}
\usepackage{rotating}
\usepackage{pdflscape}

\usepackage[dvipsnames,svgnames,x11names]{xcolor}

\begin{document}

\begin{frontmatter}
\title{Flexible multivariate spatio-temporal Hawkes process models of terrorism}
\runtitle{Multivariate spatio-temporal Hawkes process}

\begin{aug}
\author[A]{\fnms{Mikyoung} \snm{Jun}\ead[label=e1]{mjun@central.uh.edu}}
\and
\author[B]{\fnms{Scott} \snm{Cook}\ead[label=e2]{sjcook@tamu.edu}}

\address[A]{Department of Mathematics,
University of Houston,
\printead{e1}}

\address[B]{Department of Political Science,
Texas A\&M University,
\printead{e2}}
\end{aug}

\begin{abstract} 
We develop flexible multivariate spatio-temporal Hawkes process models to analyze patterns of terrorism. Previous applications of point process methods to political violence data mainly utilize temporal Hawkes process models, neglecting spatial variation in these attack patterns. This limits what can be learned from these models, as any effective counter-terrorism strategy requires knowledge on both when and where attacks are likely to occur. Even the existing work on spatio-temporal Hawkes processes imposes restrictions on the triggering function that are not well-suited for terrorism data. Therefore, we generalize the structure of the spatio-temporal triggering function considerably, allowing for nonseparability, nonstationarity, and cross-triggering (across multiple terror groups). To demonstrate the utility of our models, we analyze two samples of real-world terrorism data: Afghanistan (2002-2013) as a univariate analysis and Nigeria (2009-2017) as a bivariate analysis. Jointly, these two studies demonstrate that our generalized models outperform standard Hawkes process models, besting widely-used alternatives in overall model fit and revealing spatio-temporal patterns that are, by construction, masked in these models (e.g., increasing dispersion in cross-triggering over time). 
\end{abstract}


\end{frontmatter}

\section{Introduction}
\label{sec:intro}

Terrorism -- that is, ``the premeditated use or threat to use violence by individuals or subnational groups to obtain a political or social objective'' \cite[][pg. 4]{enders2006distribution} -- has been a focus of analytical research for at least four decades, generating research in economics \citep{schneider2015economics}, political science \citep{sandler2014analytical}, psychology \citep{crenshaw2000psychology}, sociology \citep{turk2004sociology}, and statistics \citep{tench2016spatio}. Terrorist events produce thousands of casualties annually, injure countless more \citep{stein1999medical}, and have a lasting psychological \citep{rubin2007enduring} and economic \citep{sandler2008economic} impacts. 
 
Modeling the location and timing of terrorism, however, remains a complex problem as attack location and timing are purposefully selected to evade easy prediction. Updating strategies, tactics, and behaviors throughout the course of a terror campaign can produce complex spatio-temporal patterns of attacks. Moreover, the observed attack patterns are often a consequence of actions by multiple actors (e.g., multiple terror groups, state agents, etc.), where interdependence between attack patterns by these actors is possible. Existing research indicates a contagion-like pattern in terrorism, whereby initial attacks make future events more likely are nearby locations \citep{polo2020terrorism, Siebeneck2009}. When present, such interdependence can produce terror patterns that would not be expected from analyzing a single terror group in isolation or aggregate totals across groups. As such, a model allowing both complex spatio-temporal dynamics \emph{and} interactions between multiple terror groups is essential to analyze terrorism.
 
To better model these dynamics, we develop flexible multivariate spatio-temporal Hawkes process models. Our proposed method extends existing spatio-temporal Hawkes process models in three ways. First, our spatio-temporal triggering functions are non-separable in the spatio-temporal domain, which more flexibly captures spatio-temporal characteristics contained in these datasets. Second, since many spatio-temporal point pattern datasets from real applications are not stationary or isotropic, we allow the triggering functions to be nonstationary. Finally, we introduce a flexible cross-triggering structure to account for potential interactions in multivariate spatio-temporal point patterns. Each of these additions to the triggering structure is given in a parametric form and defined with a moderate number of parameters. 
 
Using data from the Global Terrorism Database \citep{gtd}, we demonstrate the gains from our model in two studies of terrorism. In the first, we analyze 3,170 terror attacks by the Taliban in Afghanistan between 2002-2013. Even in this univariate setting -- i.e., analyzing terror attacks by a single perpetrator -- there are significant gains from our model, as these data demonstrate clear spatio-temporal nonstationarity in the triggering structure and background rate. Second, we analyze patterns of terrorism in Nigeria between 2010 and 2017, during which two major terror groups, Boko Haram and Fulani Extremists, carried out 2,084 and 496 attacks, respectively. Our results demonstrate that the attack patterns of these two groups follow distinct, but related, spatio-temporal patterns, as captured by our cross-triggering structure. We also observe clear evidence of spatio-temporal non-separability in these patterns. In sum, both analyses demonstrate the limitations of na\"ively applying existing Hawkes process models, and the benefits of incorporating a more general triggering function. The importance of flexibility in the triggering structure is shown to be especially useful in contexts, like terrorism studies, where covariate data are missing or otherwise limited.
 
The outline of the paper is as follows. In Section 2, we briefly survey the existing literature on spatial, temporal, and spatio-temporal point process models, focusing primarily on Hawkes processes. Section 3 introduces our application data on patterns of terrorism in Afghanistan and Nigeria. In Section 4, we detail the proposed multivariate spatio-temporal Hawkes process model, describing both the model features and estimation strategy. In Section 5, we demonstrate how our spatio-temporal Hawkes process model offers gains over existing methods in analyzing terror attack patterns in Afghanistan and Nigeria. Finally, we conclude by summarizing our results and noting avenues for future research in Section 6. 

\section{Background}

While there is an extensive literature on the determinants of terrorism in political science, economics, and statistics, the vast majority of this research focuses on the frequency of terror events in highly aggregated spatial units (often countries). Only recently, have researchers begun to evaluate the local-level determinants of terrorism  \citep{findley2012terrorism,nemeth2014primacy, marineau2020local}. Even here, however, researchers often use high resolution grid data -- e.g., The Peace Research Institute Oslo (PRIO) data grid cells with 0.5 by 0.5 degree dimensions -- rather than undertake point process analysis directly. Given that the available data have specific geospatial locations (e.g., latitude-longitude), aggregating these data is unnecessarily limiting \citep{lin_et_al21}, constraining the flexibility of possible models and risking bias from the well-known Modifiable Areal Unit Problem (MAUP). 

A few notable exceptions analyze terrorism data as points \citep[e.g.,][]{python_et_al19, python_et_al19_JRSSA, lin_et_al21,wang}, however, even these papers use quite restrictive models. \cite{python_et_al19}, for example, uses a Bayesian hierarchical modeling approach for spatial and temporal point data in an analysis of global terrorism. Instead of using point process models,  they apply a logistic regression model to these data, thereby modeling the probability of an event. For the error field of the logistic regression, they used a spatio-temporal separable covariance structure. In a related article, \cite{python_et_al19_JRSSA} use a similar model to \cite{python_et_al19}, employing R-INLA for computation, to again analyze global terrorism data. More recently, \cite{lin_et_al21} apply a spatial point process model, the {\it Log-Gaussian Cox Process (LGCP)}, to bivariate terrorism data in Nigeria. While LGCP models are relatively easy to implement, they may not be able to naturally capture the so-called {\it self-exciting} nature of terrorism data. 

In the existing statistics literature more broadly, several point process models have been used for related ``presence only'' point process data. In particular, LGCP models and Hawkes process models are popular \citep{schoenberg03,diggle_et_al13,jun_et_al19}. Hawkes process models, in particular, are well suited for a variety of applications, including earthquake frequency, disease growth, and conflict events. The standard self-exciting temporal Hawkes process has a conditional intensity function, i.e., given all past events up to time $t$, $\mathcal{H}_t$,
\begin{align}
\lambda(t|\mathcal{H}_t) = \mu(t) + \sum_{i:t_i < t} g(t,t_i).\label{temp_intensity}
\end{align}
Here, $\mu(t)$ is the background rate that is often modeled as a constant. The function $g$ is so-called {\it triggering function}, and  $\{t_1,\ldots,t_n\}$ denote the observed sequence of  times of the $n$  events. The triggering function, $g$, can be essentially any function, as long as $\lambda$ is non-negative with a non-negative background rate $\mu$. Unlike spatio-temporal covariance functions, the triggering function does \emph{not} need to be positive definite. Recent work has extended temporal Hawkes process models in several areas. For example, \cite{chen_hall16} utilize non-parametric estimation of intensity function for self-exciting temporal point process models. \cite{le18} consider a multivariate Hawkes process framework where observations have time gaps in the temporal domain. To model infectious disease, \cite{schoenberg_et_al19} introduce a recursive variant of the univariate, temporal Hawkes process, where the conditional intensity at a given time depends on the intensity in prior time points.

Temporal Hawkes process models have also been used in the analysis of political violence. \cite{porter2012self} use point process models to study the daily number of terrorist attacks in Indonesia from 1994 to 2007. Specifically, they employ {\it hurdle} models that consist of two parts: 1) a Bernoulli distribution to model the presence (or absence) of a terrorist event and 2) a shot noise process to model the number of terrorist attacks conditional on the presence of a terrorist event. \cite{porter2012self} focus exclusively on the timing of attacks, with no information on the location, the attack type, the group responsible, etc. Similarly, \cite{white_et_al13} analyze terrorism patterns in three Southeast Asian countries -- in separate models -- using Hawkes process models that only consider the timing of attacks. \cite{tench2016spatio} model point patterns of improvised explosive device (IED) attacks in Northern Ireland during ``The Troubles'' using a multivariate Hawkes process model, yet it is only defined in the temporal domain. \cite{johnson_et_al18} analyze the daily count of conflict events in South America using temporal self-exciting point process models. Finally, \cite{mohler_et_al20} develop a temporal Hawkes process framework to model the process of conflict events that trigger tweets. 

As such, these studies neglect spatial information contained in these data, which is necessary if researchers hope to understand where terror events are likely to occur. Move over, spatio-temporal models enable researchers to model the spatial processes that govern terror attacks patterns. The literature on interstate terrorism, for example, provide substantial evidence of contagion-like effects \citep{bove2016does, braithwaite2007transnational, neumayer2010galton}. \citet[][pg. 309]{neumayer2010galton}, for example, argue that ``[i]f a terrorist group conducts a successful terrorist attack, it will more likely attempt to launch similar attacks in the future\ldots  terrorist attacks are contagious, this time leading to spatial dependence.'' While there is less work on domestic, subnational terrorism, there is evidence to suggest that similar learning dynamics govern these events \citep{polo2020terrorism, Siebeneck2009}. Polo (2020, pg. 1917) offers a general theory to explain spillovers of domestic terrorism, arguing that ``groups observe and emulate the tactical choice of others whom they perceive as similar to them and as an example for their own behavior.'' These spatial dynamics are similar to those found in cognate applications (e.g., disease outbreak, crime, etc.) where researchers have applied \textit{spatial} and \textit{spatio-temporal} point process models using the Hawkes process framework. 

In the most standard form, a spatio-temporal Hawkes model is defined on the planar spatial domain for event locations, $\mathbf{s}\in D \subset \mathbb{R}^2$, across time, $t \in [0,T)$, and a history of spatio-temporal point patterns up to time $t$, $\mathcal{H}^{\mathbf{s}}_t$ (superscript $\mathbf{s}$ used to differentiate this from the previous $\mathcal{H}_t$; $\mathcal{H}_t$ contains only time information, whereas $\mathcal{H}^{\mathbf{s}}_t$ contains both spatial location and time information), with the conditional intensity function given as
\begin{align*}
\lambda(\mathbf{s},t|\mathcal{H}^s_t) = \lim_{\Delta \mathbf{s}, \Delta t \rightarrow 0} \displaystyle \frac{E [N\{ B(\mathbf{s}, \Delta \mathbf{s}) \times [t, t+\Delta t)\} | \mathcal{H}^{\mathbf{s}}_t   ] }{|B(\mathbf{s}, \Delta \mathbf{s})|\Delta t}. 
\end{align*}
\noindent Here, $N(A)$ is the counting measure of events over the set $A \subset D \times [0,T)$ and $|B(\mathbf{s}, \Delta \mathbf{s})|$ is the Lebesgue measure of the ball $B(\mathbf{s}, \Delta \mathbf{s})$ with radius $\Delta \mathbf{s}$ (thus, $\Delta \mathbf{s}$ is a scalar). The related conditional intensity for a \emph{self-exciting} Hawkes spatio-temporal point process is 
\begin{align}
\lambda(\mathbf{s},t|\mathcal{H}^{\mathbf{s}}_t) = \mu(\mathbf{s},t) + \sum_{i:t_i < t} g(\mathbf{s},t,\mathbf{s}_i,t_i),\label{intensity}
\end{align}
where $\mu$ is the background rate of events, $g$ is the spatio-temporal triggering function, and  $\{\mathbf{s}_1,\ldots,\mathbf{s}_n\}$ and $\{t_1,\ldots,t_n\}$ denote the observed sequence of locations and times of $n$ events.  

Our survey of this literature confirms that even here, researchers utilize simplifying assumptions (e.g., separable triggering functions) that we aim to generalize beyond. \cite{mohler_et_al11} adapt self-exciting point process models commonly used in seismology and earthquake studies to model crime. However, they consider spatio-temporal self-exciting point process models with a triggering function that is {\it separable} in spatio-temporal domain -- i.e., the spatio-temporal triggering function is factorized into spatial and temporal triggering functions. Even the spatial component of the triggering function is separable along the two spatial dimensions, which is a clear limitation for real-world applications. \cite{schoenberg16} apply spatio-temporal Hawkes models to earthquake data in Southern California, with temperature included as part of conditional intensity function. However, as in \cite{mohler_et_al11}, \cite{schoenberg16} assume a separable spatio-temporal structure with the process. Recent work on spatio-temporal Hawkes models continues this trend of assuming a separable spatio-temporal structure in the triggering function \citep{cheng_et_al18,liu_et_al20,reinhart18}.

The limitations of assuming a separable spatio-temporal structure in the presence of spatio-temporal interactions has received close attention in the geostatistical literature, where there has been extensive research on nonseparable spatio-temporal covariance models \citep[e.g.,][]{gneiting,stein05,jun_stein07}. To our knowledge, however, there has been little effort to develop non-separable spatio-temporal Hawkes process models. For example, in their comprehensive review on spatio-temporal point process modeling,
\cite{gonzalez_et_al16} acknowledge that the separability assumption is quite restrictive and note that little work has rigorously addressed the issue of separability. In addition to spatio-temporal separability, it is also common to define spatio-temporal point process models using stationary (in fact, isotropic) intensity functions in the existing literature. This is certainly true for Hawkes process case, as the intensity and triggering function depend on space and time only through spatial and temporal lags \citep{reinhart18}. As far as the authors are aware, most nonstationary Hawkes process models have the time/space/space-time dependent $\mu$ term in \eqref{intensity}, but their triggering functions are in stationary or often isotropic forms \citep[e.g.][]{chen_hall13}.

\section{Case selection}
\label{sec:data}

We focus on two countries that suffer from frequent terrorist attacks: Afghanistan and Nigeria. Each is regularly among the top five countries most affected by terrorism annually. In 2018, for example, Afghanistan and Nigeria had the first (7,379) and second-most (2,040) terror-related deaths respectively \citep{gti2020}. That these are both countries where deadly attacks occur with regularity, unfortunately, makes them ideal cases here for both substantive and statistical reasons.

Using data from the Global Terrorism Database (GTD), we separately analyze both of these country samples \citep{gtd}. The GTD data are recorded at the incident (i.e., terror attack) level, with date (at the daily level) and location (latitude and longitude) information reported for each event. As such, these data are similar to other ``presence-only'' data regularly found in population ecology and related areas, where spatio-temporal point process models have already seen wide use \citep{renner2015point}. 

With terrorism data, however, one key limitation is that political violence tends to be the most frequent in areas where we have the least robust covariate data. As such, in our preliminary analysis, we focus on two variables for which we have reliable high-resolution data: elevation and population. Population data (counts) is obtained from the WorldPop project website, {\it https://www.worldpop.org}. The spatial resolution of the population data we used is originally at 1 km spatial resolution. Because Afghanistan has not had a detailed census in decades (most recently in 1979), these population values are estimates drawn from survey research, collaboration with international organizations, and, ultimately, predictive modeling. While these are the best available sub-national data on population in Afghanistan that we are aware of, it remains likely that there is some error in these estimates. As such, this is an issue we return to later in Section 5 when discussing model specifications and results. Elevation data are obtained from an {\sf R} package, {\it elevatr} \citep{elevatr}, which pulls the elevation data from the U.S. Geological Survey (USGS) elevation point query service. We pulled elevation data for the same grid of the population data considered; spatial grid resolution we used for computation is $\mbox{0.06}^{\circ} \times \mbox{0.05}^{\circ}$ (for longitude $\times$ latitude). 

These data help us to illustrate the apparent differences in the attack patterns observed between the two countries and their correlates. A comprehensive analysis of these patterns, however, is reserved for Section~\ref{sec:analysis}.   

\subsection{Afghanistan}

In the 21st century, Afghanistan has consistently had amongst the highest incidents of terror by any measure, and in 2020 it ranked No. 1 globally according to the Global Terrorism Index (GTI) \citep{gti2020}. Over this time, there has been one main actor responsible for these attacks: the Taliban. From 2002-2013, for example, of the 5,796 terrorist attacks in Afghanistan recorded in the Global Terrorism Database \citep{gtd}, 3,179 were carried out by the Taliban (about 55\%). As recently as 2019, the Taliban received the ignominious designation as the ``world's deadliest terrorist group'' by the GTI  \citep{gti2020}. While the Taliban have recently (August 2021) assumed control of the government in Afghanistan, they remain closely associated with acts of political repression and terror. More importantly for our analysis, the Taliban were active participants in terrorism throughout the sample period (2002 to 2013).

While there has been little formal statistical analysis of these attacks of the type we consider here, there are existing spatial analyses we can draw from. \cite{fuhriman_et_al17}, for example, studied terrorist events in Afghanistan from 2002 to 2013, and found that elevation was a key factor in the spatial pattern for terror attacks over this period. While the  \cite{fuhriman_et_al17} analysis is merely suggestive, it offers a useful point of entry for our analysis.  

\begin{figure}[bt!]\centering
\begin{subfigure}[b]{0.5\textwidth}
\centering
 \includegraphics[width=\textwidth]{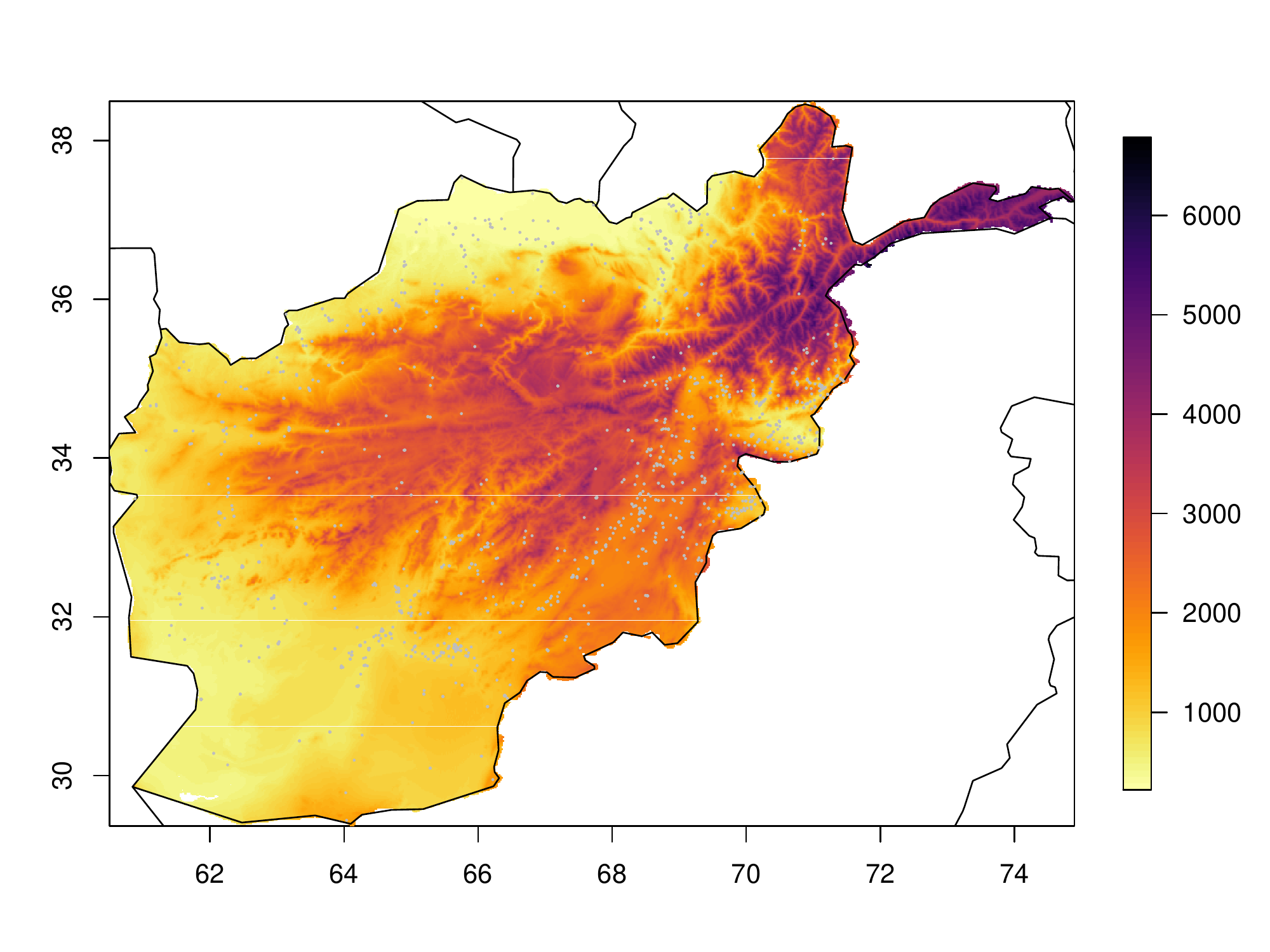}
\caption{}
\label{afg3}
\end{subfigure}
\hfill
\begin{subfigure}[b]{0.49\textwidth}
\centering
 \includegraphics[width=\textwidth]{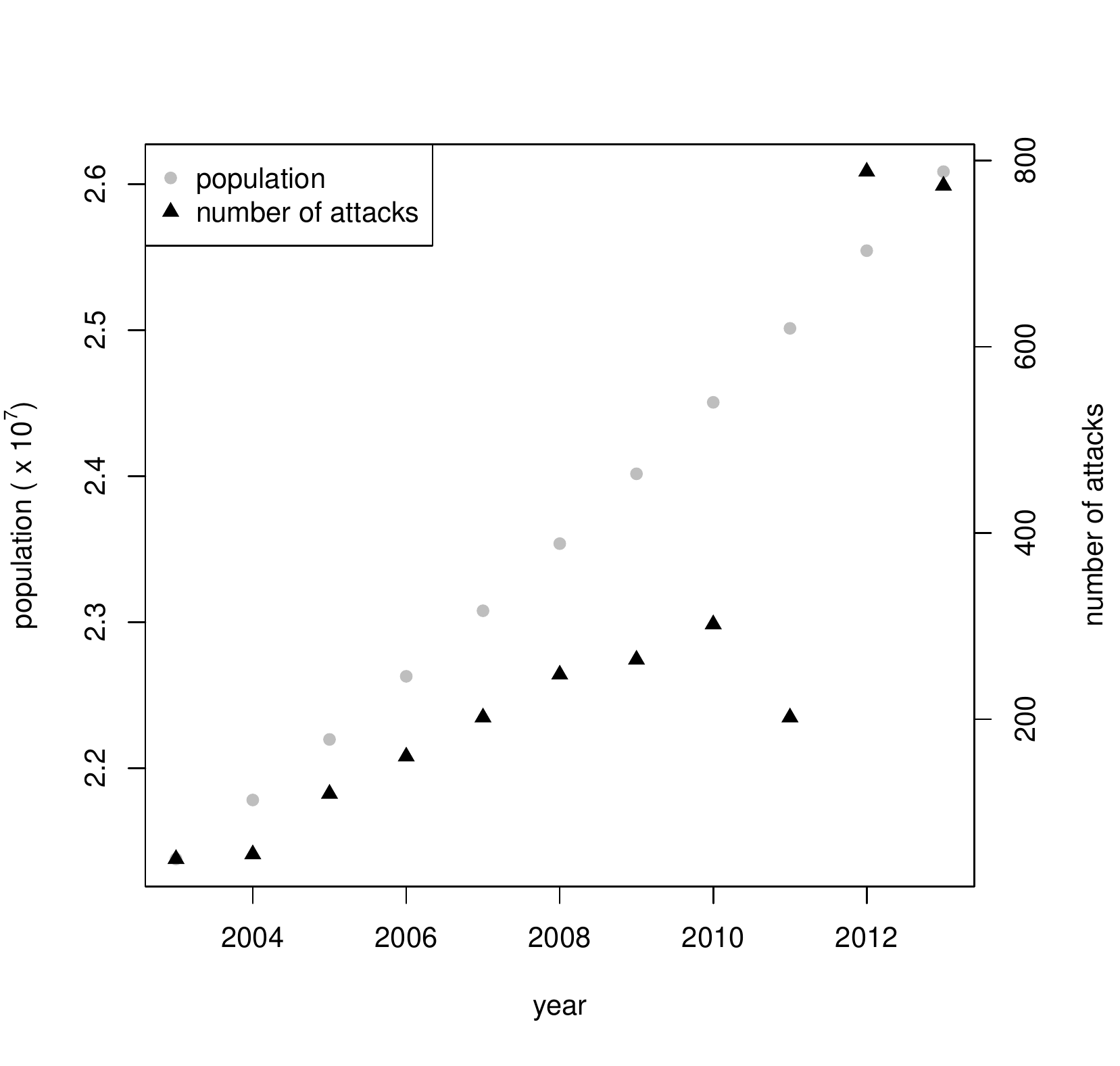}
\caption{}
\label{afg4}
\end{subfigure}
\caption{(a) Location of attacks by Taliban during 2002-2013 (marked with gray dots) and elevation (unit: meter) (b) Population and annual total number of attacks by Taliban in Afghanistan.}
\label{afg}
\end{figure}

As in \cite{fuhriman_et_al17}, we consider data from 2002 to 2013. Figure~\ref{afg3} shows the spatial pattern of attacks by the Taliban over the entire study period, with elevation represented in the background, as this was the main focus of \cite{fuhriman_et_al17}. As reported in \cite{fuhriman_et_al17}, we observe an apparent association between elevation and terrorism attacks: attacks seem to be less frequent in areas with very low or very high elevation. However, most attacks happen near National Highway 1, a so-called ``ring road'' (or circular road) that spans this area, connecting multiple major cities. Roughly speaking, this highway is in the elevation range of around 2,000 meters and there are rarely attacks in very low or high elevation areas. Therefore, if we were to incorporate elevation as a covariate in our spatial and/or spatio-temporal models, the relationship between attack patterns and the elevation variable need to be somewhat complex: 1) the effect of elevation on attacks is non-linear (roughly quadratic), 2) the effect of elevation appears to just proxy for distance to roadways and/or cities. 

\begin{figure}[hbt!]\centering
   { 
     { {\includegraphics[angle=0,totalheight=10.5cm]{ 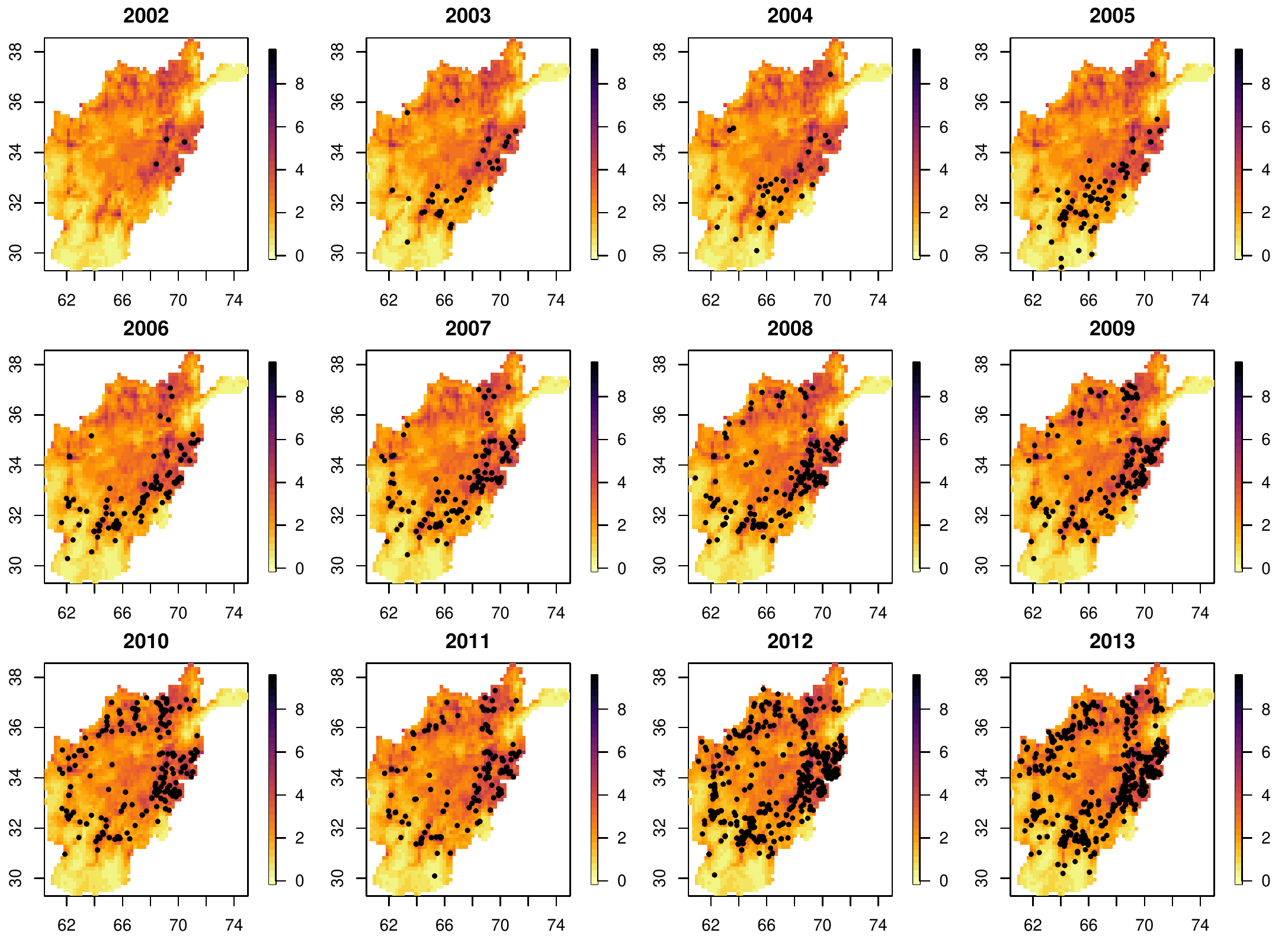}}}}
\caption{Spatial point patterns of attacks by Taliban (denoted by black dots) per year with log population in the background.}
\label{afg1}
\end{figure}

A limitation of the \cite{fuhriman_et_al17} analysis is that it fails to meaningfully consider over-time variation in the spatial distribution of these terror attacks, instead pooling all attacks across the sample period. Among other things, this limits our ability to understand how location choices may have evolved over the course of an extended terror campaign. Figure~\ref{afg4} shows the number of daily attacks over time perpetrated by the Taliban. Overall, the number of attacks increase sharply over time, indicating a clear positive trend. Moreover, in Figure~\ref{afg4} we include population as a determinant of terror attacks, since previous literature has argued that population is, and see an apparent positive correlation here as well. 

To further consider the spatio-temporal relationship between terror attacks and population, we plot the locations of attacks per year against logged population in Figure~\ref{afg1}. We observe a strong association between annual spatial patterns of attacks and population. Readers will notice that points (i.e., the terror attacks) are most frequent near large cities. This also agrees with the temporal relationship between number of attacks and population, as shown in Figure~\ref{afg4}. Therefore, we find that log of population may be a more effective covariate to consider in describing terror patterns in Afghanistan than elevation itself. This is true despite the possible measurement error in the supplied high-resolution population data. Importantly, this demonstrates the need for careful spatial \emph{and} temporal modeling of terrorism, as the spatial patterns clearly vary across time in Figure~\ref{afg1}. Simply focusing on inputs that vary only spatially (e.g., elevation) or only temporally (e.g., linear time trend) would miss these \emph{spatio-temporal} dynamics.

\begin{figure}[hbt!]\centering
   { 
     { {\includegraphics[angle=0,totalheight=5cm]{ 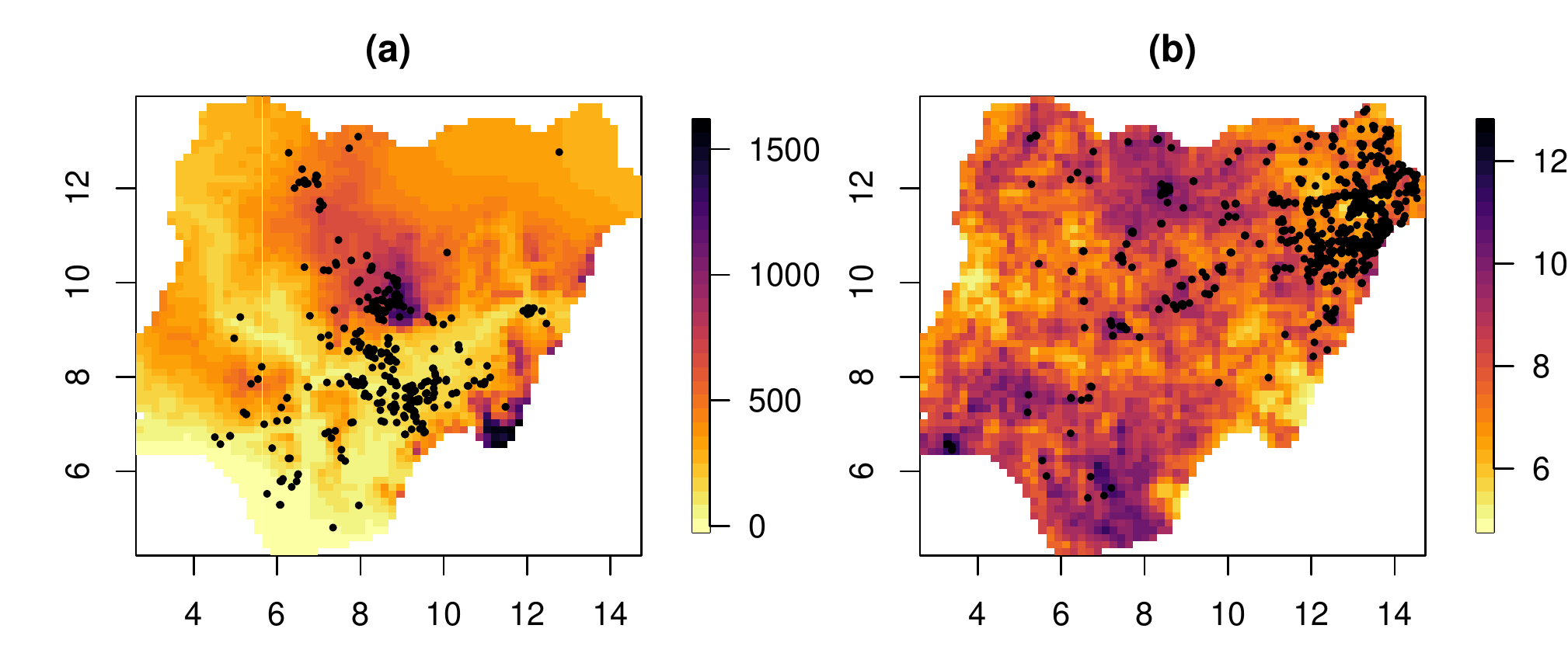}}}}
\caption{(a) elevation (unit: meter) (b) log-transformed population in 2014. In addition, (a) shows locations of attacks by Fulani extremists and (b) shows those by Boko Haram as black dots. }
\label{nigeria-elev-pop}
\end{figure}

\subsection{Nigeria}

Nigeria has also experienced high rates of political violence in recent decades. Unlike Afghanistan, there were two main groups, Boko Haram (BH) and Fulani Extremists (FE), responsible for the vast majority of these attacks, with each engaging in hundreds of attacks during our sample period, 2009 to 2017. Specifically, there were 2,075 attacks and 493 attacks, carried out by BH and FE, respectively. While a univariate analysis suffices for the Afghanistan sample, the presence of two active terror groups in Nigeria allows us to further consider the bivariate structure required to model these attack patterns jointly. As we can see in Figure~\ref{nigeria-elev-pop}, the spatial distribution of terror attacks (in 2014) for the two groups vary in a number of ways -- e.g., FE attacks tend to occur in the center of the country (the farm belt), whereas BH attacks tend to occur in the northeast. Unlike the Afghanistan analysis, neither elevation nor population immediately seem to be strongly associated with these terror attack patterns in our preliminary analysis. As Figure~\ref{nigeria-elev-pop} shows, the spatial patterns of attacks by either group do not align well with the spatial patterns of elevation or population.

\begin{figure}[hbt!]
\centering
\begin{subfigure}[b]{0.9\textwidth}
\centering
\vspace{-2mm}
\includegraphics[angle=0,totalheight=5cm]{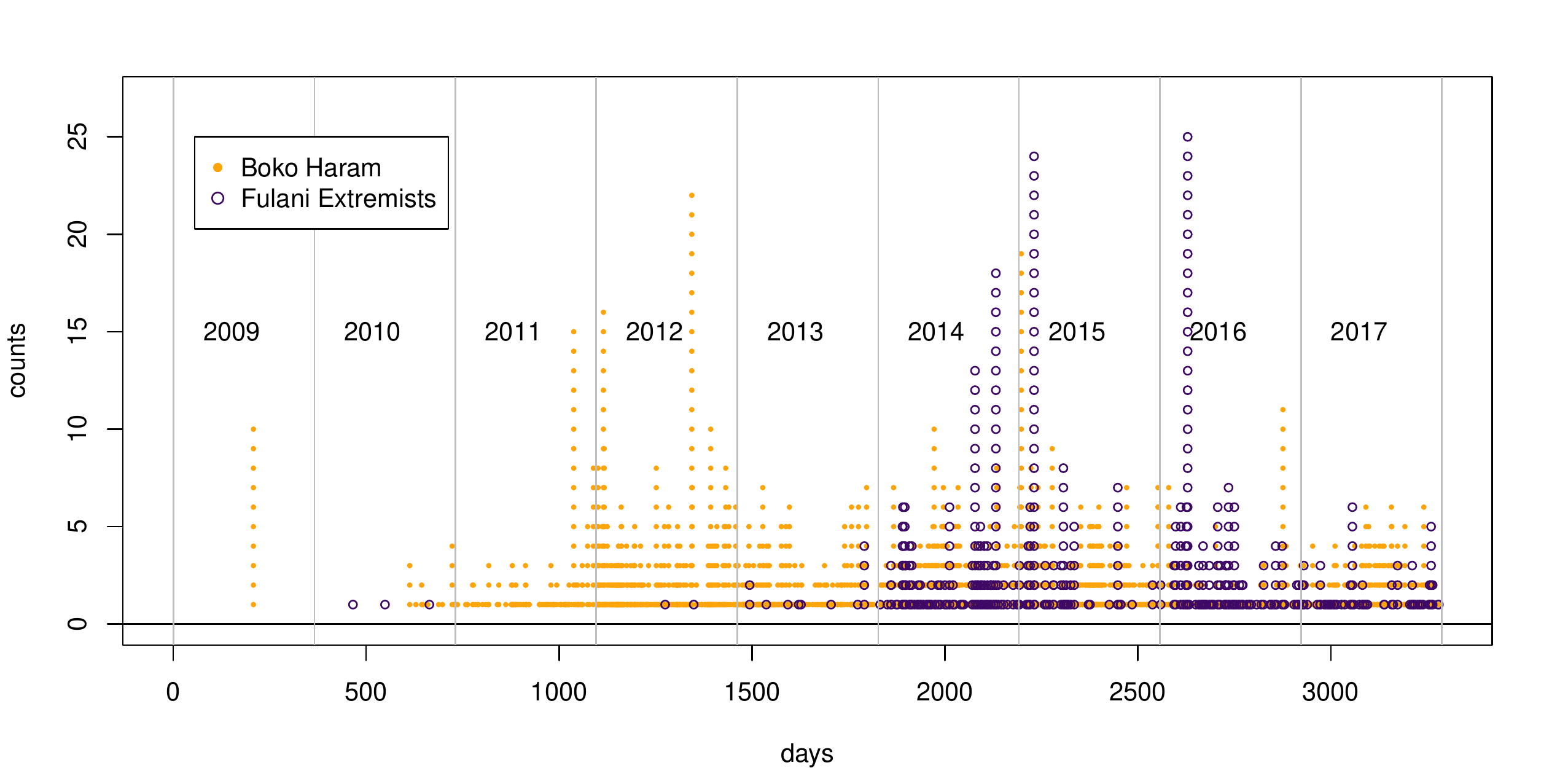}
\caption{Daily counts of terrorism in Nigeria.}
\label{counts-new}
\end{subfigure}
\vspace{5mm}
\begin{subfigure}[b]{0.9\textwidth}
\centering
\vspace{-2mm}
 \includegraphics[angle=0,totalheight=5cm]{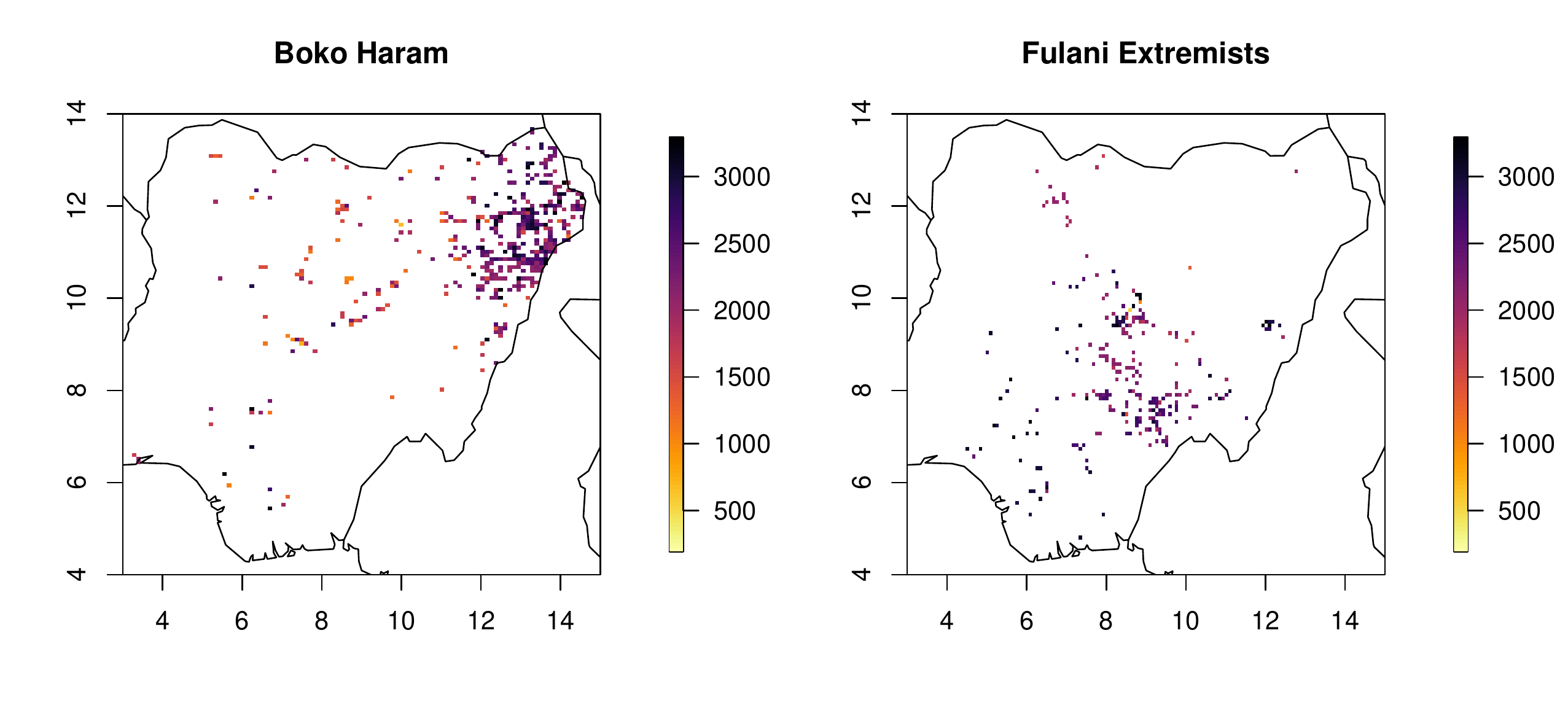}
\caption{Spatial patterns of terrorism in Nigeria. The color scale represents the time of attacks relative to the first day in the sample (yellow is earlier, purple is later).}
\label{spatial-new}
\end{subfigure}
\caption{Terror attacks by Boko Haram and Fulani Extremists, 2009-2017. }
\label{nigeria-data}
\end{figure}

To consider variation across time, Figure~\ref{counts-new} shows number of attacks over time between 2009 and 2017 by the two groups. While we observe that Boko Haram was much more active than Fulani extremists early on in our sample, there is no other obvious temporal pattern apparent in these counts. Moreover, from this visual alone, it does not appear that there is an over-time relationship between the two groups when space is neglected. Given that, we next consider variation in time \emph{and} space in  Figure~\ref{spatial-new}, which aggregates all attacks between 2009 and 2017 for each of the two groups and plots these spatially. The color scale shows the time of attacks (days relative to the first day in the time period, i.e. January 1, 2009). As before, attacks by Boko Haram are concentrated in the northeast corner of the country, while those by Fulani Extremists are mainly in the center of the country. More interestingly, it appears that the attacks by Boko Haram become increasingly concentrated in the northeast corner of Nigeria later in the sample period. This demonstrates the need for spatio-temporal modeling of these patterns, as this variation would be lost in models looking exclusively at time or space. 

\subsection{Understanding spatio-temporal and bivariate structure}
\label{sec:empirical}

Ultimately, a richer understanding of these terror patterns requires more than simple visualizations. As such, we now further explore the spatio-temporal patterns of these terror attacks in both countries. This helps to determine the spatio-temporal structure of the triggering functions, both marginal and joint, used in the Hawkes process models below.

\begin{figure}[hbt!]\centering
   { 
     { {\includegraphics[angle=0,totalheight=5cm]{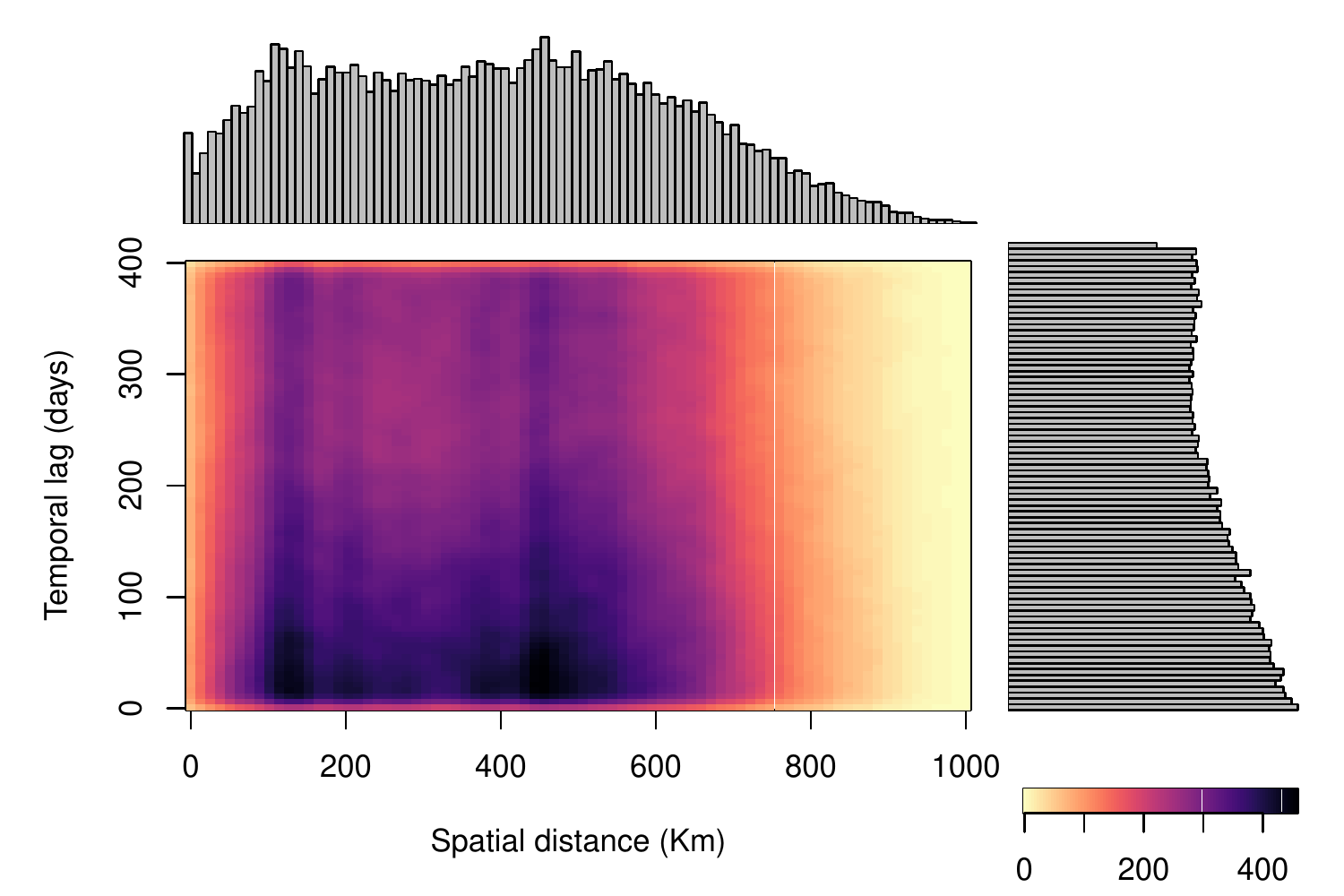}}}}
\caption{Two-dimensional histogram of spatial and temporal lags between pairs of attacks by Taliban, 2002-2013.}
\label{afg-counts}
\end{figure}

Figure~\ref{afg-counts} provides histogram(s) of counts of \emph{pairs} of attacks by the Taliban plotted against the spatial and temporal lags (i.e., distance and time between the pair of events). Consider, for example, a single attack (recorded in space and time), we calculate the pairwise distance (i.e., the spatial lag) and time (i.e., the temporal lag) between this observation (the attack) and all previous attacks in the sample. To formalize this some, let  $T_i=\{T_{i,1},\ldots,T_{i,n(i)}\}$ denote a collection of locations of attacks by the Taliban at time $i$, and $n(i)$ reflect the number of attacks at time $i$ -- e.g., if $n(i)=0$, then $T_i$ is a null set. Further, let $d(T_i, T_j)$ denote the distance matrix (of size $n(i) \times n(j)$) between the locations of attacks at time $i$ and time $j$ (where $i > j$). Figure~\ref{afg-counts} then shows binned averages of the distribution of elements of $d(T_i,T_j)$ and $i-j$. For instance, for two time points $i$ and $j$, where $i > j$ and $n(i), n(j) \neq 0$, there are $n(i) \times n(j)$ pairs of attacks, and a corresponding number of time lags (i.e., $i-j$). Similarly, for this time lag $i-j$, we calculate spatial distance for $n(i) \times n(j)$ pairs of attacks, with the spatial distances reported as the spatial lags in Figure~\ref{afg-counts}. Here we display up to 400 days for the time lag and up to 1,000 km for spatial lag, with 80 bins for each dimension. 

As expected, in Figure~\ref{afg-counts} we see that counts of pairs generally decay as either the spatial or temporal lags increase. It also appears that the shapes of histogram of spatial lags may have subtle changes with temporal lags (for instance, horizontal slices of histogram for temporal lag 0 to 100 days look somewhat different from those for temporal lag 200 to 400 days), which may indicate signs of spatio-temporal non-separability. However, such changes may also be due to spatially and/or temporally varying background rates, so consider various specifications for the background structure in Section~\ref{sec:analysis} when analyzing these data.

We calculate these empirical quantities for the Nigerian case as well using similar notation, with minor difference given the bivariate structure of these data. That is, $B_i$ is a collection of location of attacks by Boko Haram at time $i$ and $F_i$ by Fulani Extremists. Let $m(i)$ and $l(i)$ denote the number of attacks at time $i$ (i.e. number of spatial locations for the attacks) by Boko Haram and Fulani Extremists, respectively. For the interaction between the two groups, given $B_i$'s and $F_j$'s, let $d(B_i, F_j)$ denote the distance matrix between the locations of attacks by Boko Haram and Fulani Extremists at time $i$ and time $j$. These distance matrices are only defined when there is at least one attack by each group to be considered.  

\begin{figure}[hbt!]
\begin{subfigure}{.49\textwidth}
  \centering
  \includegraphics[width=.9\linewidth]{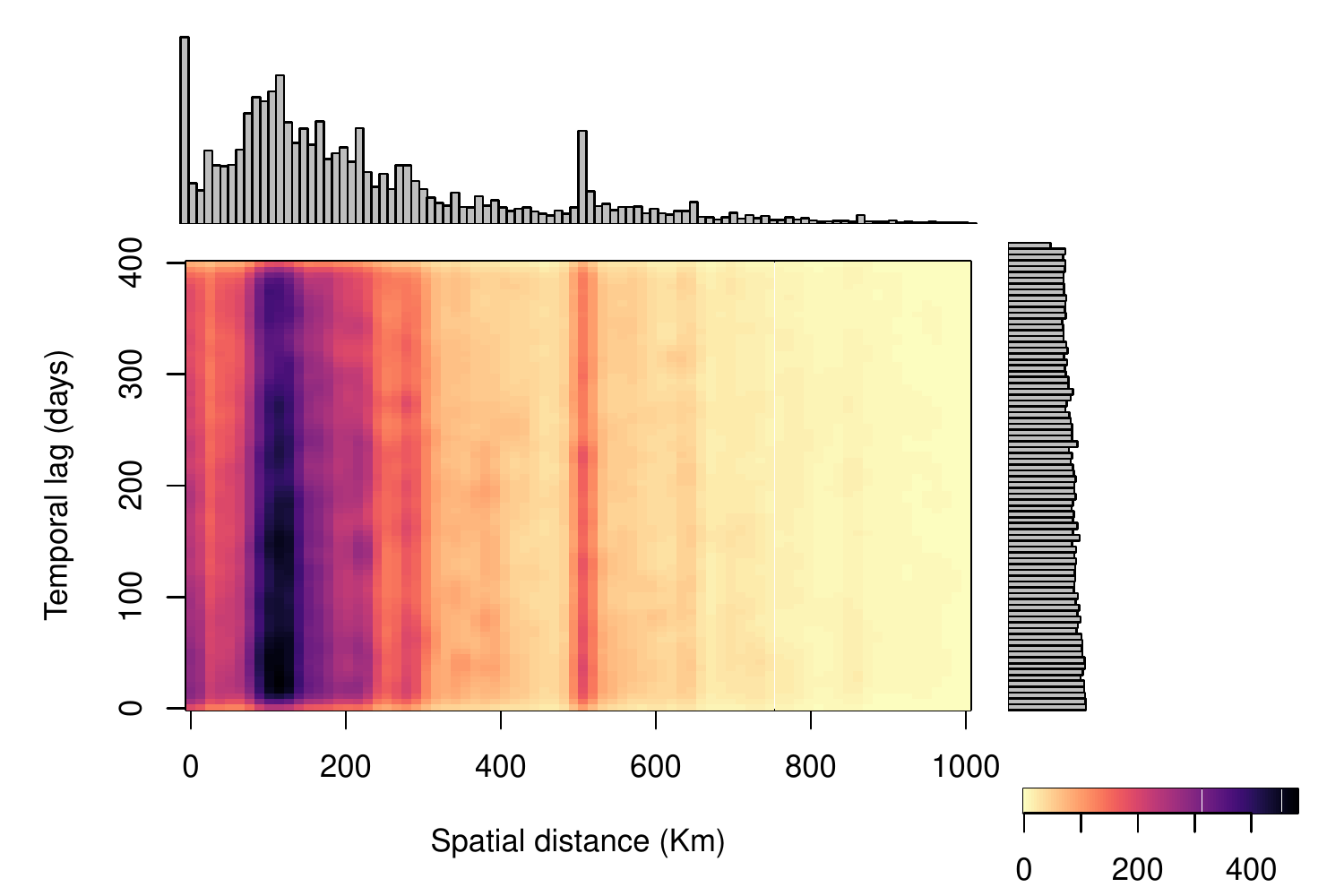}  
  \caption{BH}
  \label{fig:sub-first}
\end{subfigure}
\begin{subfigure}{.49\textwidth}
  \centering
  \includegraphics[width=.9\linewidth]{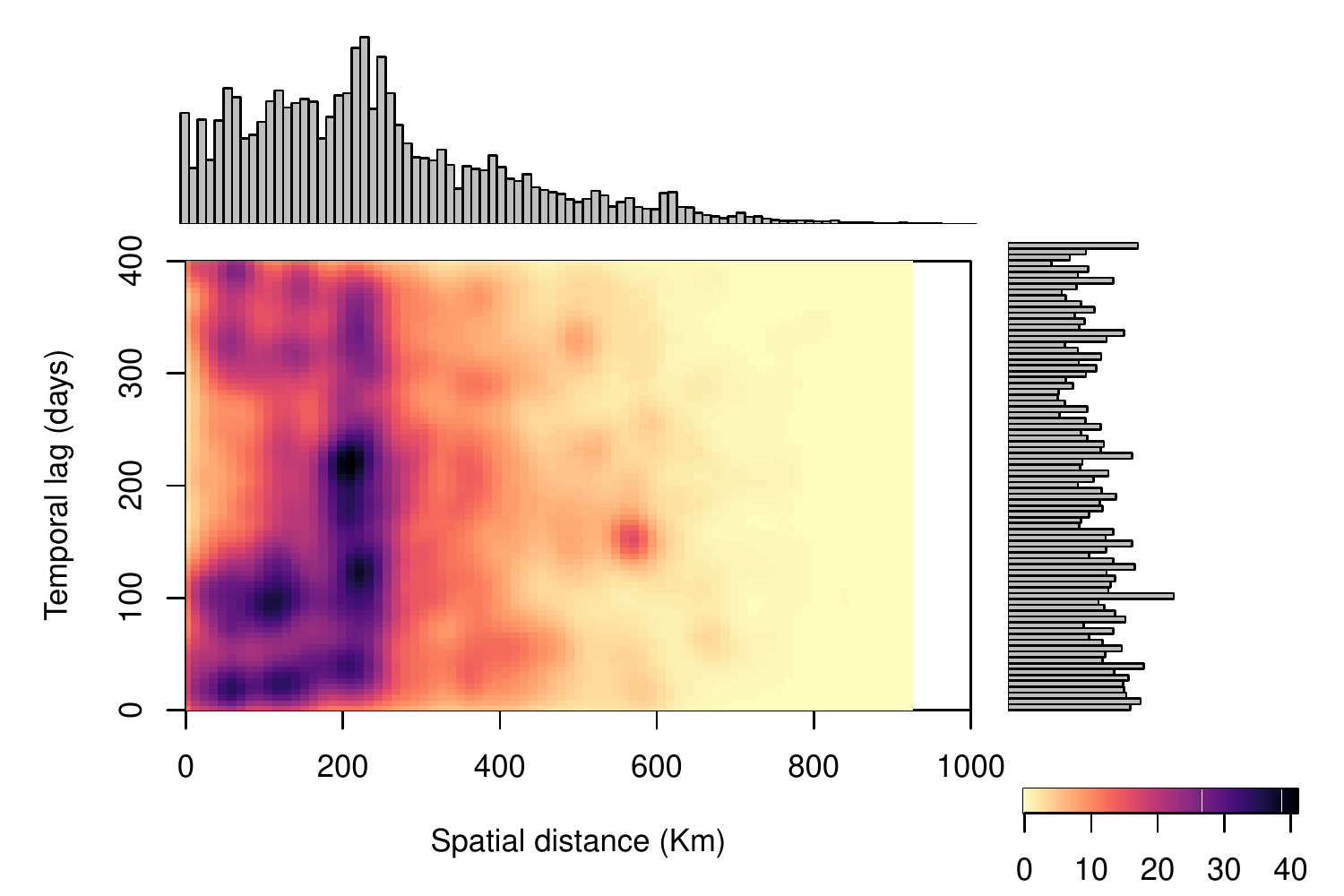}  
  \caption{FE}
  \label{fig:sub-second}
\end{subfigure}
\begin{subfigure}{.49\textwidth}
  \centering
  \includegraphics[width=.9\linewidth]{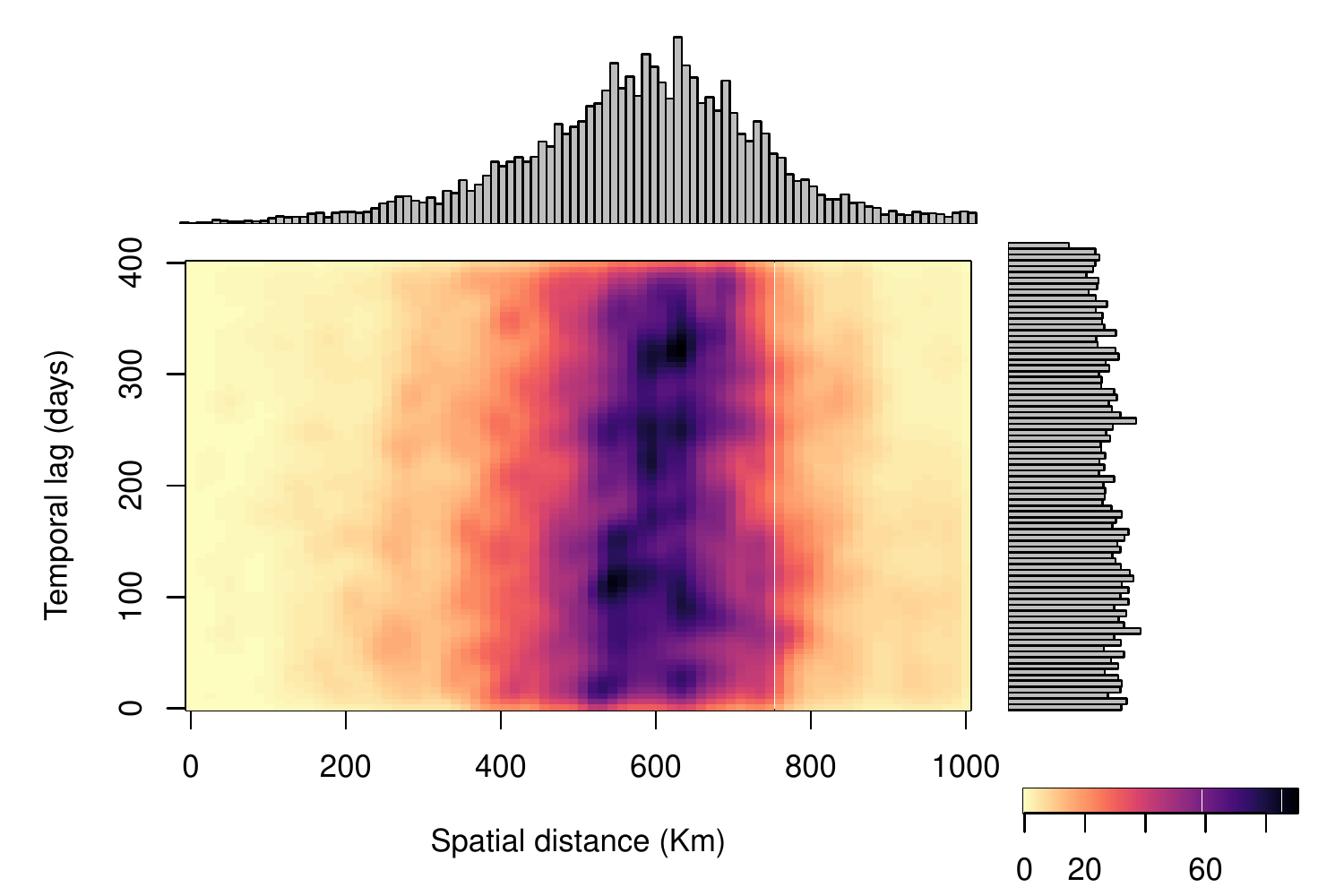}  
  \caption{BH(FE)}
  \label{fig:sub-second}
\end{subfigure}
\begin{subfigure}{.49\textwidth}
  \centering
  \includegraphics[width=.9\linewidth]{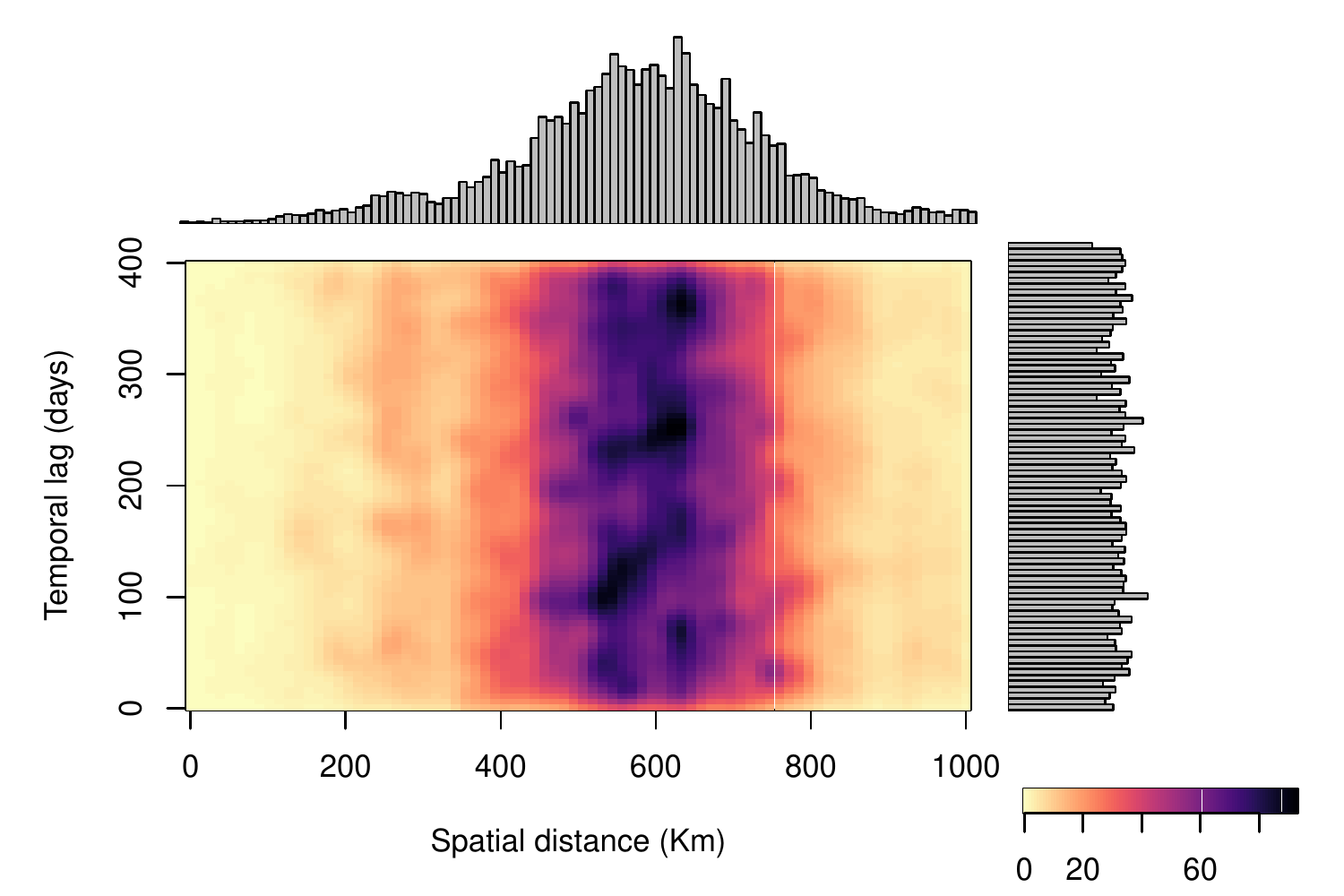}  
  \caption{FE(BH)}
  \label{fig:sub-second}
\end{subfigure}
\caption{Two-dimensional histogram of spatial and temporal lags between pairs of attacks for Nigeria.}
\label{nig-spacetime}
\end{figure}

In the remaining discussion, we will use ``BH'' to denote Boko Haram, and ``FE''  to denote Fulani Extremists. Therefore, BH(FE) denotes pairs of events where an attack by FE was followed by an attack by BH (i.e., FE-then-BH), and FE(BH) denoting the alternative sequence (i.e., BH-then-FE). Figure~\ref{nig-spacetime} is similar to Figure~\ref{afg-counts}, but now separately given for BH (a) and FE (b). The cross-group interactions are also reported in Figure~\ref{nig-spacetime}, with BH(FE) given in (c) and FE(BH) given in (d). 

Several interesting relationships seem apparent from Figure~\ref{nig-spacetime}. First, these results suggest that the triggering functions (marginal and cross) decay slowly across time if we assume the background rate is constant in space and time. Second, we see that for FE there were no two attacks were more (spatially) distant than roughly 920 km (within temporal lag values up to 400 days). Third, spatio-temporal non-separability for marginal FE in panel (b) is more apparent than marginal BH in panel (a), assuming constant background rate. Fourth, in panels (c) and (d) we see that the mode of the distribution for the spatial lags between two attacks by different groups is around 600 km. This is unlike the marginal cases, where the number of pairs generally decreases as the spatial lag increases. Importantly, this suggests that the functional form of the spatial triggering function needs to be different for the marginal and cross cases, which is not commonly done in the existing literature.

We now leverage this preliminary analysis to identify necessary generalizations to the Hawkes process model in the next section, and then apply these more flexible models to these terrorism data in Section \ref{sec:analysis}.

\section{Flexible spatio-temporal Hawkes process models}

As shown in Section~\ref{sec:empirical}, the terror attack patterns in Afghanistan and Nigeria exhibit complex, yet distinct, spatial and temporal characteristics. Furthermore, the attack patterns for the two terror groups in Nigeria seem to indicate complex \emph{interactions} between different terror groups that cannot be dealt with using many existing point process models. Given this, we develop multivariate spatio-temporal Hawkes process models suitable for data exhibiting these properties. In particular, we propose generalizing the standard Hawkes process model to permit nonstationary and spatio-temporally non-separable triggering functions. Additionally, we develop multivariate Hawkes process models that can describe interactions between multiple spatial patterns (e.g., attacks from multiple terror groups). 


While not the main focus of our analysis, we also consider different specifications of the background rate, $\mu$ in \eqref{intensity}. Initially, we use a constant background rate, focusing exclusively on the specification of the triggering function. However, in spatial and spatio-temporal Hawkes process models the parameters in the triggering function can proxy for an under specified background rate, so we also consider more general specifications of the background structure. As we detail below, we allow for both spatially and temporally varying background rates modeled as a function of appropriate covariates. As we illustrate in Section~\ref{sec:analysis}, a spatially and temporally varying background function is useful in analyzing the Afghanistan data, but did not matter as significantly when analyzing the Nigeria data. Across all specifications of the background rate, we continue to see evidence that a more general triggering structure, of the type we develop here, offers gains in analyzing these data.
 

 \subsection{Spatio-temporal triggering functions}
 
Consider the general structure of the (conditional) intensity function for spatio-temporal Hawkes processes given in \eqref{intensity}. Drawing on our preliminary analysis of terrorism attack patterns in Section~\ref{sec:empirical}, we extend upon this base function and introduce flexible representations of spatio-temporal triggering function $g$ in both univariate and multivariate settings. 

We start with a separable space-time structure for $g$ (as in \cite{reinhart_greenhouse18}): 
\begin{align}
g^{[1]}(\mathbf{s},t,\mathbf{w},u;\alpha,\beta,\phi)= \alpha \cdot \Bigl \{\frac{1}{\beta} \exp\Bigl(-\frac{t-u}{\beta}\Bigr)\Bigr \}\cdot \Bigl\{ \frac{1}{{2 \pi} \phi^2}  \exp\Bigl(-\frac{|\mathbf{s}-\mathbf{w}|^2}{2 \phi^2}\Bigr)\Bigr\}.   \label{separable1}
\end{align} 
Here, $(\mathbf{s}, t)$ and $(\mathbf{w}, u)$ are two spatio-temporal ``locations" where the spatio-temporal triggering function is defined ($t > u$). As the temporal and spatial components of the function $g^{[1]}$ are density functions (that integrate to 1), $\alpha$ is a parameter that determines the level of spatio-temporal triggering. Parameters $\beta$ and $\phi$ determine the temporal and spatial length scale of triggering, respectively.

Note that $g^{[1]}$ is separable in space and time (as long as $\alpha$, $\beta$, and $\phi$ are constants) and each of the spatial and temporal components are density functions. For cases like our sample of terror events in Afghanistan, where a nonstationary spatio-temporal triggering structure may be beneficial, we extend $g^{[1]}$ in two ways, that is,  $$g^{[1]}(\mathbf{s},t,\mathbf{w},u;\tilde{\alpha},\beta,\phi),$$ or $$g^{[1]}(\mathbf{s},t,\mathbf{w},u;{\alpha},\beta,\tilde{\phi}),$$ with $\tilde{\alpha}=\tilde{\alpha}(\mathbf{s}, t, \mathbf{w}, u)$ and $\tilde{\phi}=\tilde{\phi}(\mathbf{s}, t, \mathbf{w}, u)$ as functions that depend on spatial locations and time points (cf. $\alpha$, $\beta$, and $\phi$ are constants). Both versions of the spatio-temporal triggering function are motivated by a nonstationary extension of isotropic covariance functions, allowing either the variance or spatial range parameter to vary over space and time \citep[e.g.,][]{stein05_cises,jun11}. 

In our analysis of the Afghanistan terror data, we use the log-transformed population variable (specifically, log of population plus 1, to avoid the problem for a spatial pixel and time with zero population) to model $\tilde{\alpha}$ and $\tilde{\phi}$. Specifically, we model $\tilde{\alpha}$ and $\tilde{\phi}$, respectively, as a function of $lP(\mathbf{w},t)$ -- the log transformed population at location $\mathbf{w}$ and time $t$ divided by the maximum log population value over the entire sample period (ensuring $lP \leq 1$) -- as: 
\begin{align} \tilde{\alpha}(\mathbf{s},t,\mathbf{w},u) = \alpha_0 \cdot \displaystyle   \frac{lP(\mathbf{s},t)+lP(\mathbf{w},u)}{2},
\label{alpha}
\end{align} and 
\begin{align}\tilde{\phi}(\mathbf{s},t,\mathbf{w},u) =\phi_0 +\phi_1 \cdot\displaystyle  \frac{lP(\mathbf{s},t)+lP(\mathbf{w},u)}{2},
\label{phi}
\end{align} with constants $\alpha_0$, $\phi_0>0$. Furthermore, we set $\phi_1 > -  \phi_0$ so the resulting $\tilde{\phi}$ in \eqref{phi} is strictly positive, while $\tilde{\phi}$ can be positively or negatively associated with the population through $\phi_1$ (as $\phi_1$ can be positive, zero, or negative). For \eqref{alpha}, the resulting triggering structure implies that the level of triggering is increasing in population, that is, more intense triggering for higher population values, which is consistent with the observed pattern of terror data we reported in Section~\ref{sec:data}. Note that both \eqref{alpha} and \eqref{phi} produce a non-separable spatio-temporal triggering structure, as each depends on pairs of spatial locations and temporal points in a non-separable way. This is still somewhat limiting in many cases, such as our Nigerian sample of terror events, which may be better modeled using a more direct non-separable spatio-temporal triggering function.

With our Nigerian case in mind, we now focus on incorporating a flexible cross-triggering structure as well as spatio-temporal nonseparability. In bivariate Hawkes processes, the triggering function $g$ becomes a $2 \times 2$ matrix, $$ g =\displaystyle \begin{pmatrix}g_{kk} & g_{kl}\\ g_{lk} & g_{ll}\end{pmatrix},$$ where $k$ indicates one sample of points and $l$ the other (ex. Boko Haram and Fulani Extremists). The diagonal elements, $g_{kk}$ and $g_{ll}$, are the marginal triggering function for $k$\textsuperscript{th} and $l$\textsuperscript{th} point patterns, respectively, and the off-diagonal elements, $g_{kl}$ and $g_{lk}$, are the {\it cross-triggering}, that is, an event by a group (the column index) triggered by events in the other group (the row index). Note that we allow the model to have $g_{kl} \neq g_{lk}$ for $k \neq l$. 

In much of the literature on bivariate and multivariate Hawkes process models, the same spatial and/or spatio-temporal structure is used in both the marginal and cross-triggering functions \citep{yuan_et_al19}. Generally, the level of triggering decreases in spatial distance (i.e., spatial lags) and/or time (i.e., temporal lags). {The multivariate Hawkes process models proposed in \cite{pmlr-v162-soliman22a} also induce ``mutual" triggering. }In our Nigeria example, however, such cross-triggering function may not be suitable given the clear spatial \emph{separation} across the two groups. As our preliminary analysis above suggests, we should consider spatial clustering in the marginal processes but spatial repulsion in the joint process. However, we are not aware of Hawkes process models that enable this in the existing literature.

\begin{figure}[hbt!]\centering
   { 
     { {\includegraphics[angle=0,totalheight=5cm]{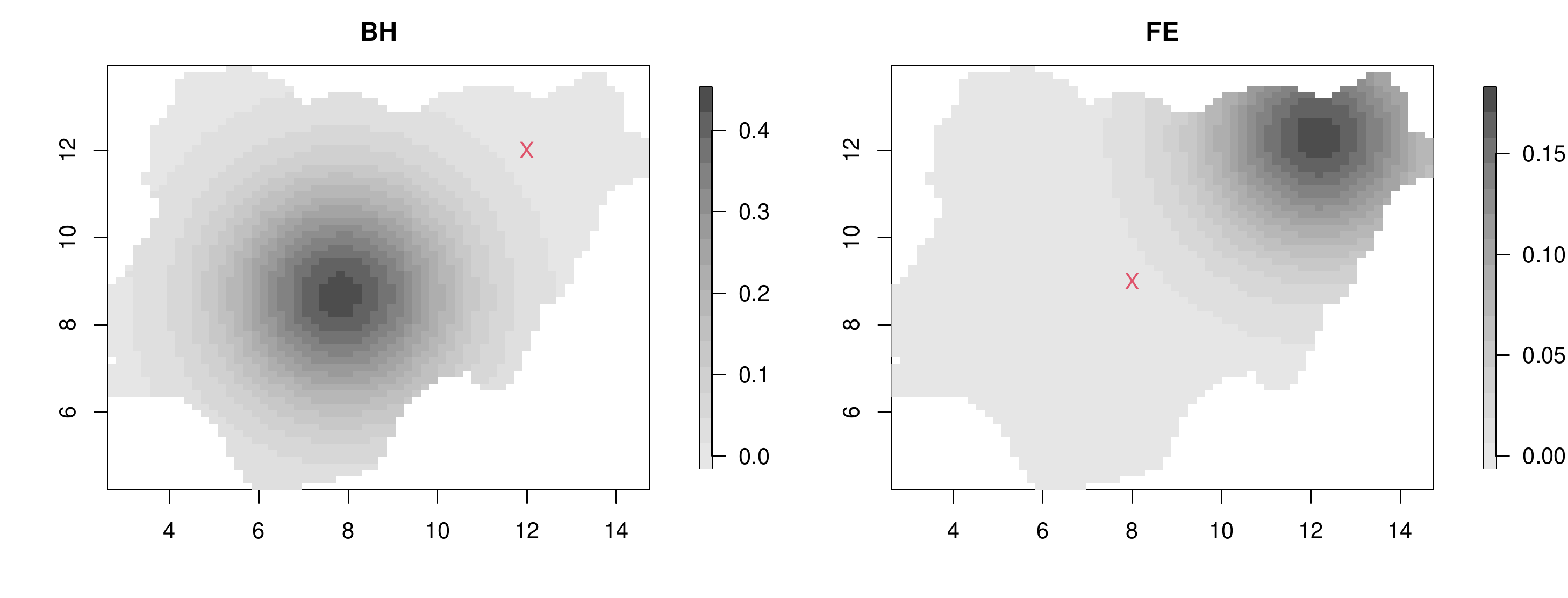}}}}
\caption{Fitted cross-triggering function for the Nigerian data with estimated parameter values of M2-6 presented in Section~\ref{sec:result-nigeria}. The left panel shows fitted triggering effect for attacks by FE, triggered by an attack by BH (location marked in red cross). The right panel instead shows fitted triggering effect for attacks by BH, triggered by an attack by FE (location marked in red cross). }
\label{nig-cross}
\end{figure}

Therefore, we consider the following structure for the spatio-temporal cross-triggering function,
\begin{align}
g^{[2]}(\mathbf{s},t,\mathbf{w},u;\alpha,\beta,\phi,\mathbf{m}&)=\label{cross}\\ &\alpha \cdot \Bigl \{\frac{1}{\beta} \exp\Bigl(-\frac{t-u}{\beta}\Bigr)\Bigr \}\cdot \Bigl[\frac{1}{{2 \pi^2} \phi}  \exp\Bigl\{-\frac{|(\mathbf{s}-\mathbf{w})-\mathbf{m}|^2}{2 \phi^2}\Bigr\}\Bigr], \nonumber
\end{align}
and set $g_{kl}=g^{[2]}$ for $k \neq l$ ($t>u)$. Here the key is the extra parameter vector, $\mathbf{m} \in \mathbb{R}^2$, which allows spatial separation between the two spatial patterns. That is, the spatial component of the cross-triggering function in \eqref{cross} is not a monotonically decreasing function of spatial distance. Instead, it increases up to a certain spatial lag, which induces spatial separation between the two spatial point patterns. Figure~\ref{nig-cross} illustrates this point. As is shown, the triggering of events for the other group is not a monotonic function of the spatial lag. With \eqref{cross}, we can estimate the maximum spatial separation distance determined by fitted \textbf{m}=($\eta$, $\xi$) along with other parameters. As we demonstrate later in Section~\ref{sec:result-nigeria}, we find that adding \textbf{m} is effective in bivariate modeling of terrorism data for Nigeria. Note that the spatial component in \eqref{cross} is a two-dimensional (Gaussian) density function. 

Finally, we also explore spatio-temporal nonseparability with a simple nonseparable spatio-temporal triggering functions. For instance, adapting ideas in \cite{gneiting} for nonseparable spatio-temporal nonseparable covariance functions, we set 
\begin{align}
g^{[3]}(\mathbf{s},t,\mathbf{w},u;\alpha,\beta,\phi,\mathbf{m},\gamma)=&\alpha \cdot \Bigl \{\frac{1}{\beta} \exp\Bigl(-\frac{t-u}{\beta}\Bigr)\Bigr \} \label{nonseparable1}\\  & \times \Bigl[\frac{1}{{2 \pi} \phi^2(1+(t-u)/\beta)^{\gamma}}  \exp\Bigl\{-\frac{|(\mathbf{s}-\mathbf{w})-\mathbf{m}|^2}{2 \phi^2(1+(t-u)/\beta)^{\gamma}}\Bigr\}\Bigr], \nonumber
\end{align}
with $0<\gamma<1$.

Having now defined $g^{[1]}$, $g^{[2]}$, $g^{[3]}$, and their refinements, we can summarize all the implied models that we apply in our terrorism analysis. For the Afghanistan data, we explore different versions of the background rate, including a constant background function and spatio-temporally varying background functions: for $\mu$ in \eqref{intensity}, $$ \mu(\mathbf{s}, t) = \beta_0 + \beta_1 X(\mathbf{s}, t)$$ with appropriate covariate $X$. Combining these with the triggering functions above, we estimate the following set of univariate spatio-temporal Hawkes process models (triggering function; background rate):
\begin{itemize}
\item[][M1-1] Poisson process (i.e., no triggering); Standardized log population used for $X$ in $\mu$,
\item[] [M1-2] Poisson process (i.e., no triggering); Standardized time used for $X$ in $\mu$,
\item[] [M1-3] Nonstationary model with $g = g^{[1]}$ with constant $\alpha$ and $\tilde{\phi}$ as in \eqref{phi}; Standardized log population used for $X$ in $\mu$ (as in M1-1),
\item[][M1-4] Nonstationary model with $g = g^{[1]}$ with constant $\alpha$ and $\tilde{\phi}$ as in \eqref{phi}; Set $\beta_1=0$ in $\mu$ (i.e., constant background rate),
\item[][M1-5] Isotropic model with $g=g^{[1]}$ but with a Gaussian (as opposed to exponential) temporal triggering function; Standardized log population used for $X$ in $\mu$ (as in M1-1). 
\end{itemize}

For the Nigerian terrorism data, we also allowed the background rate $\mu$ to vary spatially (or spatio-temporally), but the fits of these specifications were noticeably worse than those models with a constant background rate. This is discussed in more detail in Section 5.2. As such, the model variations we describe here focus on variation in the triggering function. Specifically, we consider an extensive set of models to explore the bivariate structure and spatio-temporal non-separability: 
\begin{itemize}
\item[][M2-1] Isotropic univariate Hawkes process model fitted to a single point pattern (merged Boko Haram and Fulani Extremist attacks) with triggering function given by $g^{[1]}$,
      \item[][M2-2] Bivariate model with \{$g_{kk},g_{ll}\}$ given by $g^{[1]}$ and $g_{kl} = g_{lk}=0$ for $k \neq l$. That is, the two patterns are fitted separately, but there is no cross-triggering in the model, 
    \item[][M2-3] Bivariate model with \{$g_{kk},g_{kl},g_{lk},g_{ll}\}$ given by $g^{[1]}$, and $g_{kl} = g_{lk}$. Note that the parameters for $g_{kk}$ and $g_{ll}$ are freely estimated. 
    \item[][M2-4] Bivariate model with $\{g_{kk}, g_{ll}\}$ given by $g^{[1]}$ and $\{g_{kl}, g_{lk}\}$ by $g^{[2]}$ ($g_{kl}$ and $g_{lk}$ have a common $\mathbf{m}=(\eta_c, \xi_c)$), with $
      g_{kl}=g_{lk}=g^{[2]}(\cdot,\cdot,\cdot,\cdot;\cdot,\cdot,\cdot,\mathbf{m})$ for $k \neq l$.
     \item[][M2-5] Same as M2-4 except that we set $g_{kl}=g^{[2]}(\cdot,\cdot,\cdot,\cdot;\cdot,\cdot,\cdot,\mathbf{m})$ and \\$g_{lk}=g^{[2]}(\cdot,\cdot,\cdot,\cdot;\cdot,\cdot,\cdot,-\mathbf{m}),$ for $k \neq l$.
\item[][M2-6] Bivariate model with $\{g_{kk}, g_{ll}\}$ given by  $g^{[3]}$ (with $\mathbf{m}=\mathbf{0}$), $\gamma_l$ and $\gamma_k$ are freely estimated) and  $\{g_{kl}, g_{lk}\}$ by $g^{[3]}$. Regarding $\mathbf{m}$, cross-triggering functions are set in the same way as in M2-5.
\end{itemize}
To aid readers, Table~\ref{nig-model-summary} summarizes the salient properties for the 6 models considered in our analysis of the Nigerian data. Here, ``cross-triggering'' indicates interactions across different groups in the bivariate structure, and ``non-decreasing'' reflects the fact that the triggering function is not a monotonically decreasing function of spatial lag, as with $\mathbf{m} \neq \mathbf{0}$ for $g^{[j]}$, $j=2,3$. Non-decreasing functions were used for the cross-triggering functions in M2-4, M2-5, and M2-6. For all M2-x models, exponential functions are used for temporal triggering structure, and Gaussian functions for spatial triggering structure. 

\begin{table}[bth!]
\centering
\caption{Properties of triggering functions used for bivariate analysis of Nigeria data}
\begin{tabular}{|l|c|c|c|c|c|c|}
\hline
&M2-1&M2-2&M2-3&M2-4&M2-5&M2-6\\
\hline
Bivariate&N&Y&Y&Y&Y&Y\\
Cross-triggering&N&N&Y&Y&Y&Y\\
Non-decreasing&N&N&N&Y&Y&Y\\
Nonseparable&N&N&N&N&N&Y\\

\hline
\end{tabular}
\label{nig-model-summary}
\end{table}

\subsection{Stability Conditions}
\label{sec:stability}

Stability conditions on some parameters of the triggering functions are required. For instance, \cite{jang_et_al19} discuss the stability condition -- i.e., the integral of the triggering function is less than 1 -- for the univariate case in order to ensure that each event generates ``less than one subsequent event'' on average (in space and time), and hence the process stays stable. \cite{roueff_sachs19} provides a more general condition for locally stationarity in the univariate case. Similarly, we need the following conditions to ensure stability: $|\alpha|<1$ and  $|\alpha_0|<1$ for $\tilde{\alpha}$. 

For multivariate Hawkes processes, \cite{bremaud_massoulie96} and \cite{Chen2016MultivariateHP} discuss similar conditions on some parameters of the marginal and cross-triggering functions. Specifically, for a stationary (and isotropic) bivariate Hawkes spatial process with a matrix valued triggering function given by $h=(h_{kl})_{k,l=1,2}$, the spectral radius of $h(0)$ needs to be less than 1. We impose similar conditions on each of the triggering functions used in our bivariate (i.e., M2-x) models. 

\subsection{Statistical Inference}
\label{sec:numerical}

Statistical inference is done through maximum likelihood estimation (MLE). Although estimation of parameters using {\it minimum contrast} methods is generally simpler and computationally efficient \citep{zhu_et_al22}, MLE gives more statistically efficient parameter estimates \citep{diggle}. Calculation of the likelihood functions for spatio-temporal Hawkes process models in the planar spatial domain, marked or unmarked, is straightforward compared to LGCP models. This is because LGCP models are doubly stochastic and one needs to integrate Poisson likelihood over its stochastic intensity functions, whereas Hawkes models do not have such a problem. If $\{(\mathbf{s}_1,t_1),\ldots, (\mathbf{s}_n,t_n)\}$ denotes a spatio-temporal point pattern in $D \times (0,T]$, the log-likelihood function is given by \citep{daley_vere-jones03,reinhart18}

\begin{align}
l(\Theta)= \sum_{i=1}^n \log\{\lambda(\mathbf{s}_i,t_i)\} -\int_0^T \int_D \lambda(\mathbf{s},t) d\mathbf{s} dt. \label{llik}
\end{align}
Here, $\Theta$ is a collection of parameters and $\lambda$ is the conditional intensity function for the Hawkes process as in \eqref{intensity}. The integral term in \eqref{llik} is often done numerically, not analytically. Numerical approximation of the double integral in \eqref{llik} is given by a double summation approximation over regular grids in spatio-temporal domain. 
 
For the Afghanistan data, we used 10,000 spatial grids and 800 time points for the approximation of integral. For bivariate analysis in the Nigerian case, we use common spatio-temporal grids for both processes. A coarser grid was used since we have a bivariate problem in this case (that results in spatial and temporal points as well as more parameters to be estimated): we used 2,800 spatial grid points and 500 time points. We tried finer spatial and temporal grid resolutions for both applications and found that the results do not change significantly, which indicates that these grid points are dense enough for a good approximation. In the GTD dataset, there are several terror events with the exact same longitude/latitude for a given time. This can cause computational problems, as the probability of having more than one event at the exact same spatial and temporal point is zero in any spatio-temporal process model. To avoid this problem, we jitter the spatial coordinates for those cases, adding random numbers drawn from a Normal distribution with mean zero and standard deviation 0.01 to the original longitude and latitude. See Section~\ref{sec:discussion} for further discussion on this issue. 

{\cite{chen_et_al18} discussed three information criteria for model selection for Hawkes process models, namely the Akaike's information criterion (AIC), the Bayesian information criterion (BIC), and the Hannan-Quinn (HQ) criterion. For a given model with the maximized log-likelihood value $\mathcal{L}$, $k$ number of parameters, and $n$ sample size, define $$\mbox{AIC} = -2 \mathcal{L} + 2k,$$ $$\mbox{BIC} = -2 \mathcal{L} + k \ln(n),$$ and $$\mbox{HQ} = -2 \mathcal{L} + 2k \ln(\ln(n)).$$ They compared these three criteria with theoretical and simulation studies, and argued that performance of each of the criteria depends on the model as well as sample size in a complex manner. \cite{reinhart18} commented that the AIC works well for small sample cases but not for large sample cases. Later, we compare models based on all three criteria.  }

\section{Results}
\label{sec:analysis}

We now present results from the application of these spatio-temporal Hawkes process models to our two cases, Afghanistan and Nigeria. All data and code necessary to reproduce our analysis are available as a Dataverse repository (link to be added upon article acceptance). 

\subsection{Results for Afghanistan}

Table~\ref{afg-estimate} reports the parameter estimates (and their asymptotic standard errors), maximized log-likelihood values, and the AIC, BIC, and HQ values. We first consider the two different structures on background rate: logged population (M1-1) and time (M1-2). Comparing the AIC values, we see that log population is substantially more effective than time ($-$27,893 vs. $-$20,609). This indicates that the variation in the spatial structure of the population variable is helpful in describing the point pattern. Turning now to how the triggering function affects model fit, we can compare M1-1 and M1-3 which differ only in that the latter introduces a nonstationary triggering structure. Here we see that the nonstationary triggering structure in M1-3 dramatically improves the fit, changing the AIC from $-$27,893 to $-$52,720. We also see that some of the effect previously attributed to logged population in M1-1 is now explained by the parameters in the triggering function, as indicated by the reduced magnitude of $\mu_1$ (8.60 to 7.57) and the larger standard errors. We do not, however, observe the same consequence for the parameters of the triggering function if we instead treat the background rate as constant (as in M1-4). While M1-3, with its spatially varying background rate, does offer some improvements relative to M1-4 these are relatively minor (in either AIC terms or comparing common coefficients across models).  We also fit models with spatially varying $\alpha$ and spatio-termpoal nonseparable triggering functions to the Afghanistan data utilizing $g^{[3]}$, but interestingly fits did not improve compared to M1-3 (spatially varying $\beta$) or separable structures given in Table~\ref{afg-estimate}. {Note that we discussed mainly regarding the AIC, but all the conclusions are consistent even with the BIC or HQ.}
 
What do these results tell us about the pattern of attacks in Afghanistan? First, the fitted results of M1-3 and M1-4 regarding nonstationary triggering structure indicate that spatial triggering distance increases with the population. As such, attacks in highly populated areas are likely to affect the likelihood of future attacks in a wider area than attacks in less populated areas. Second, we see that the estimated parameters for the background rate are fairly consistent across models with different triggering functions (M1-1, M1-3, and M1-5). However, the differences that do emerge suggest that underspecifying the triggering function---as would be the case by construction in other candidate models such as the Poisson or LGCP---causes one to overestimate the parameters of the base rate. Finally, since the inclusion of the spatio-temporal nonseparable triggering functions did not offer significant gains, we suspect that having logged population variable in the background rate takes care of much of spatio-temporal ``interaction," demonstrating the need to have a well-specified model of both the background and triggering function in models of this type.




\begin{table}[bt!]
    \centering
     \caption{Univariate spatio-temporal Hawkes process models and terrorism in Afghanistan, 2003-2012. For
reference, the model with no triggering and a constant background rate  gives a maximized log-likelihood value of 9,265.80 and AIC of $-$18,529.60. {Note sample size $n=3,170$. For each criterion, the best values are marked bold-faced.}}
      \label{afg-estimate}
  \begin{tabular}{|c|r|r|r|r|r|}
\hline
&M1-1&M1-2 &M1-3 &M1-4&M1-5\\
\hline
\hline
$\mu_0$&8.91e-2(4.10e-2)&2.00(5.76e-2)&4.99e-2(7.97e-2)&2.49(3.73e-2)&0.037(7.72e-2)\\
$\mu_1$&8.60(7.06e-2)&3.11(7.59e-2)&7.57(0.16)&\cellcolor{lightgray}&7.72(0.148)\\
\hline\hline
$\alpha$&\cellcolor{lightgray}&\cellcolor{lightgray}&1.00(0.037)&1.00(0.036) &0.85(6.04e-2)\\
\hline
$\beta$ (day)&\cellcolor{lightgray}&\cellcolor{lightgray}&116.52 (3.70)&115.02(3.59) &15.25(3.21)\\
\hline
$\phi$ (km) &\cellcolor{lightgray}&\cellcolor{lightgray}&\cellcolor{lightgray}&\cellcolor{lightgray}&13.71(9.62e-4)\\
$\phi_0$ (km)&\cellcolor{lightgray}&\cellcolor{lightgray}&$-$13.56(1.71e-2)&$-$13.60(1.80e-2) &\cellcolor{lightgray}\\
$\phi_1$ (km)&\cellcolor{lightgray}&\cellcolor{lightgray}&13.57(1.39e-2)&13.61(1.49e-2) &\cellcolor{lightgray}\\
\hline
\hline
$\#$ para&2&2&6&5&5\\
\hline
max loglik&13,948.59&10,306.63&{\bf 26,366.27}&25,641.08&26,033.77\\
\hline
 AIC &$-$27,893.18&$-$20,609.26&{\bf $-$52,720.54}&$-$51,272.16&$-$52,057.54\\
 \hline
 {BIC}&$-$27,881.06&$-$20,597.14&{\bf $-$52,684.17}&$-$51,241.85&$-$52,027.23\\
 \hline
 {HQ}&$-$27,888.83&$-$20,604.91&{\bf $-$52,707.49}&$-$51,262.29&$-$52,046.67\\
\hline
\end{tabular}
\end{table}

\begin{figure}[bth!]\centering
\begin{subfigure}[b]{0.69\textwidth}
\centering
  \includegraphics[width=0.99\textwidth]{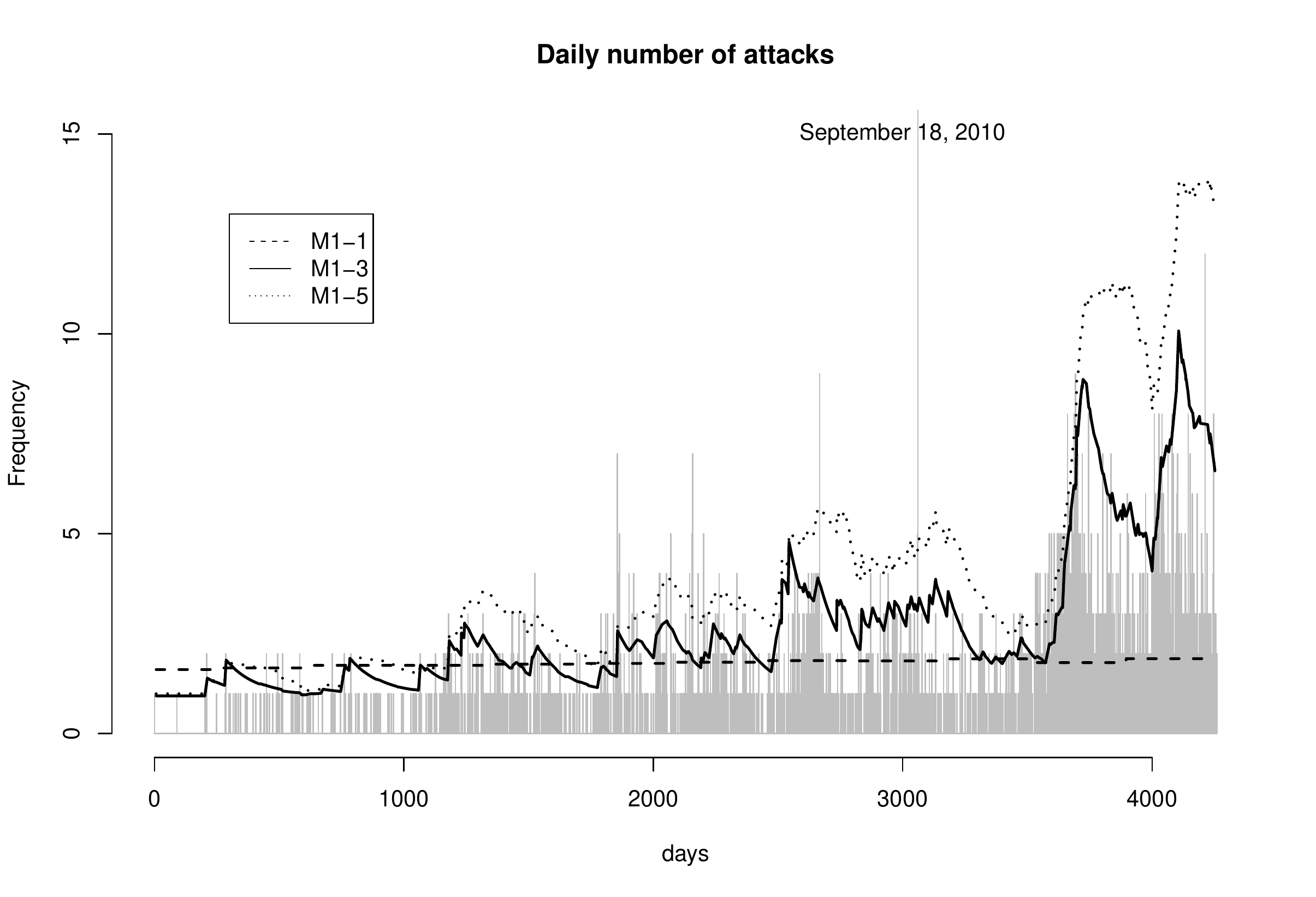}
\caption{}
\label{afg-fit-time}
\end{subfigure}
\hfill
\begin{subfigure}[b]{0.3\textwidth}
   \includegraphics[width=\textwidth]{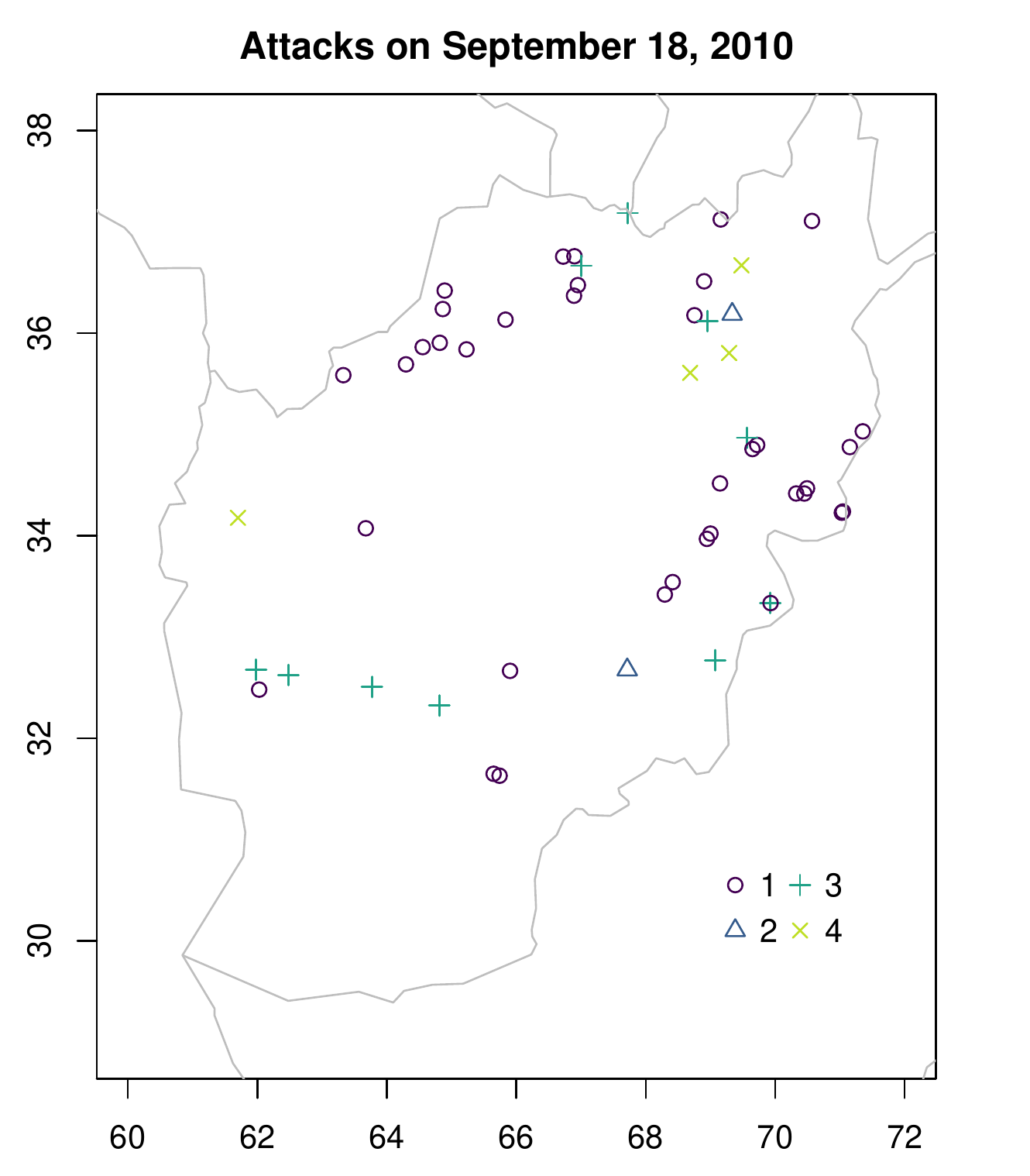}
\caption{}
\label{afg-election}
\end{subfigure}
\caption{ (a) Number of daily attacks for Afghanistan with the expected number of daily attacks from fitted spatio-temporal Hawkes process models. The maximum value for y-axis displayed is set to be 15 to show temporal patterns of fitted curves (Note: the number of attacks on September 18, 2010 is 52) (b) Location of attacks by Taliban on September 18, 2010 in Afghanistan. Each color/symbol denotes specificity value (1 to 4) -- see Section~\ref{sec:discussion} for definitions and further discussion of {\it specificity}.}
\label{afg-fit}
\end{figure}

Rather than simply compare coefficients, Figure~\ref{afg-fit-time} shows daily counts of \emph{observed} attacks by the Taliban during the sample period, and the corresponding \emph{expected} number of daily counts from three of our fitted models (M1-1, M1-3, and M1-5). Generally, the fitted curve from M1-3 best matches the overall temporal structure of the empirical counts. In contrast, the fitted curve of expected counts from M1-1 is relatively flat despite temporally-varying background function (via population). Finally, the fitted curve from M1-5 overestimates counts observed in the data and relative to the expected by M1-3. During the sample period, an obvious outlier in daily counts occurs on September 10, 2010, as there are 52 reported attacks, which is unusually large compared to the rest of the sample. That day was the 2010 Afghan Parliamentary election, and as shown in Figure~\ref{afg-election}, there were attacks by the Taliban throughout the country. This is consistent with existing evidence on the more general relationship between elections and terrorism \citep{Newman2013}. Different symbols used in Figure~\ref{afg-election} will be further discussed in Section~\ref{sec:discussion}. 

\subsection{Results for Nigeria}
\label{sec:result-nigeria}

The results from our analysis of the Nigerian terrorism data are given in Table~\ref{estimate-new-nigeria-update-new}, which includes parameter estimates (and their asymptotic standard errors), maximized log-likelihood values, and the AIC, BIC, and HQ values from our univariate and bivariate analyses. Since we use both univariate and bivariate models here, it is important to be clear about our notation to avoid confusion on specific parameters. In the univariate analysis in M2-1, the data from BH and FE are combined, so we only report common $\alpha$, $\beta$, and $\phi$ (i.e., no further indexing is required). In models M2-2 to M2-6, however, parameters that correspond to the process for BH are denoted with subscript $b$, and those for FE with $f$. Parameters for cross-triggering are denoted with subscripts $bf$ and $fb$, respectively. If particular parameters for cross-triggering are the same (for bivariate models), we denote them with subscripts $c$.  

Turning to the results, let us first note that when we fitted Poisson process models with spatially varying background rate, with covariates, longitude, latitude, and population (M2-0), results were much worse than the simplest model, M2-1, in Table~\ref{estimate-new-nigeria-update-new}. For instance, the model with longitude and latitude as linear terms in the background rate resulted in the AIC of $-$7,594.46. This value is much worse than the result of M2-1 (AIC $-$10,443.94) where the background rate is constant and a simple univariate triggering structure (using only 3 parameters) is employed. Given that, we only focus on models with a constant background rate in the rest of our discussion of the Nigerian data. 

\begin{table}[h]
  \centering
       \caption{Univariate and bivariate spatio-temporal Hawkes process models and terrorism in Nigeria, 2009-2017. For reference, the model with no triggering but the background rate linear with longitude and latitude (thus 3 parameters, same as M2-1) gives a maximized log-likelihood value of 3,800.23 and AIC of $-$7,594.46. {Note that the sample size $n=$ 2,065 (BH) $+$ 492 (FE) $=$ 2,557. For each criterion, the best values are marked bold-faced.}}
       \scalebox{0.9}{
    \begin{tabular}{|c|r|r|r|r|r|r|}
\hline
         &M2-1& M2-2&M2-3  &M2-4 &M2-5&M2-6 \\
         \hline
         \hline
           $\alpha$ &0.43(1.6e-3)&\cellcolor{lightgray}&\cellcolor{lightgray}&\cellcolor{lightgray}& \cellcolor{lightgray}&\cellcolor{lightgray}\\
         $\alpha_b$&\cellcolor{lightgray}&0.60 (1.84e-2)&0.36 (1.33e-2)&0.33 (1.26e-2)& 0.34 (1.34e-2) &0.34  (1.45e-2)\\
             $\alpha_{bf}$&\cellcolor{lightgray}&\cellcolor{lightgray}&0.71 (1.93e-3)&0.66 (1.51e-2)& 0.70 (2.11e-2)& 0.71 (2.98e-2)\\
         $\alpha_f$&\cellcolor{lightgray} &0.47 (3.13e-2)&0.19 (1.12e-2)&0.34 (1.39e-2)&  0.26 (1.15e-2)&0.24 (5.39e-2)\\
              $\alpha_{fb}$&\cellcolor{lightgray}&\cellcolor{lightgray}&0.24 (2.43e-2)&4.0e-3 (1.80e-3)& 0.29 (1.33e-2) &0.28 (2.72e-2)\\
         \hline
           $\beta$ (day) &137.83 (8.48)&\cellcolor{lightgray}&\cellcolor{lightgray}&\cellcolor{lightgray}& \cellcolor{lightgray}&\cellcolor{lightgray}\\
         $\beta_b$& \cellcolor{lightgray}&174.86 (10.87)&302.48 (18.57)&286.13 (19.64)& 283.29 (19.66)& 303.99 (25.32)\\
         $\beta_f$ &\cellcolor{lightgray}&103.24 (13.34)&317.99 (47.63)&126.02 (15.57)& 140.68 (58.27)&43.47 (7.88)\\
         $\beta_c$ &\cellcolor{lightgray}&\cellcolor{lightgray}&260.61 (30.07)&195.68 (19.58)&  201.63 (20.49)&213.15 (24.75)\\
         \hline
          $\phi$ (km) &30.78 (0.58)&\cellcolor{lightgray}&\cellcolor{lightgray}&\cellcolor{lightgray}& \cellcolor{lightgray}&\cellcolor{lightgray}\\
         $\phi_b$ &\cellcolor{lightgray}&14.31 (0.35)&1.05 (2.76e-2)&0.58 (4.07e-2)& 0.60 (3.64e-2)&1.01 (0.33)\\
         $\phi_f$ &\cellcolor{lightgray}&56.87 (3.96)&1.73 (0.17)&31.98 (2.27)& 4.50 (0.40)&4.46 (1.20)\\
         $\phi_c$  &\cellcolor{lightgray}&\cellcolor{lightgray}&528.28 (20.46)&228.61 (9.12)& 205.41 (6.63)&202.11 (6.34)\\
         \hline
         $\eta_c$ &\cellcolor{lightgray}&\cellcolor{lightgray}&\cellcolor{lightgray}&3.53 (0.16)&4.16 (0.11)&  4.16 (2.16)\\
         $\xi_c$&\cellcolor{lightgray}&\cellcolor{lightgray}&\cellcolor{lightgray}&3.21 (0.15)& 3.32 (0.11)& 3.31 (0.87)\\
         \hline
        $\gamma_b$&\cellcolor{lightgray}&\cellcolor{lightgray}&\cellcolor{lightgray}&\cellcolor{lightgray}   &\cellcolor{lightgray} &1.00 (0.11)\\
         $\gamma_f$ &\cellcolor{lightgray}&\cellcolor{lightgray}&\cellcolor{lightgray}&\cellcolor{lightgray} &\cellcolor{lightgray} &1.00 (0.11)\\
                 \hline
         \# para&3&6&10&12&12&14\\
         \hline
           loglik&5,224.97&5,391.46&8,155.75&8,378.62&8,696.67&  {\bf 8,723.65}\\
          \hline
          AIC&$-$10,443.94&$-$10,770.92&$-$16,316.53&$-$16,733.24&$-$17,369.34&{\bf $-$17,419.30}\\
          \hline
            {BIC}&$-$10,424.60&$-$10,735.84&$-$16,258.06&$-$16,663.08&$-$17,299.18&{\bf $-$17,337.45}\\
            \hline
          {HQ}&$-$10,437.58&$-$10,758.20&$-$16,295.33&$-$16,707.80&$-$17,343.90&{\bf $-$17,389.62}\\
         \hline
       \end{tabular}}
      \label{estimate-new-nigeria-update-new}
     \vspace{1ex}
     
          {\raggedright \scriptsize{  \hspace{2em} Note: $b$ indicates Boko Haram; $f$ indicates Fulani Extremists; $c$ indicates a common process; $bf$ indicates\\  \hspace{2em} non-symmetric cross-triggering (f-then-b), $fb$ indicates non-symmetric cross-triggering (b-then-f) \par}}

\end{table}

Turning to Table~\ref{estimate-new-nigeria-update-new}, we see the benefits of a less restrictive Hawkes process model. While some results are consistent across models -- i.e., patterns of attacks by Fulani Extremists exhibit larger spatial triggering range -- there are several noticeable differences across models as we introduce (or increase the complexity of) the cross-triggering function. Importantly, the models with a cross-triggering term (M2-3 to M2-6) show significant improvements in the AIC values (as compared to models without cross-triggering, such as M2-1 and M2-2). The cross-triggering, moreover, is not symmetric, i.e. $\hat{\alpha}_{bf} \neq \hat{\alpha}_{fb}$ in any of the models considered. {Similarly to the Afghanistan case, all three information criteria give consistent results in terms of model comparison.}

Even among the models that allow for cross-triggering, we see clear benefits from more flexible structure. First, comparing M2-3 with M2-4, M2-5, and M2-6 (i.e., the models that include non-monotonically decreasing cross-triggering functions), the effect of non-monotonic cross-triggering function is apparent. In comparing M2-4 and M2-5, it is reassuring to note that AIC for M2-5 was significantly smaller than that for M2-4. Note that M2-4 and M2-5 have almost the same model structure with the same number of parameters, but M2-4 does not properly take care of the interaction between the two groups. The main difference between M2-4 and M2-5 is the cross-triggering structure. The structure in M2-4 does not properly describe the interaction between the two point patterns by BH and FE. Thus, estimated $\alpha_{fb}$ is close to zero to compensate for the restriction in M2-4. On the other hand, the fitted cross-triggering function in Figure~\ref{nig-cross} (with M2-5) agrees well with the point patterns we observe in Figure~\ref{spatial-new}. 

Second, the estimated values of $\mathbf{m}$ are consistent across all models (M2-4 to M2-6). This contrasts with results from model M2-3, which does not have such a flexible cross-triggering function, where we observe an unusually large spatial triggering range, $\hat{\phi}_c = 528.28$ km (note that the longest distance in Nigeria is around 1,120 km). This result may be due to the monotonically decreasing cross-triggering function in M2-3, which cannot well describe spatial separation between the two patterns. As a result, the estimated cross-triggering spatial range is inflated to account for the two large spatial clusters (i.e., the attacks of both groups). 

Finally, comparing the model that allows for nonseparability in the cross-triggering function (M2-6), we again see gains from greater flexibility. For example, spatial-temporal nonseparability in the marginal triggering structure for both processes are evident, which does have important consequences for our understanding of the group-specific patterns.

\begin{figure}[hbt!]
\centering
\includegraphics[width=\textwidth]{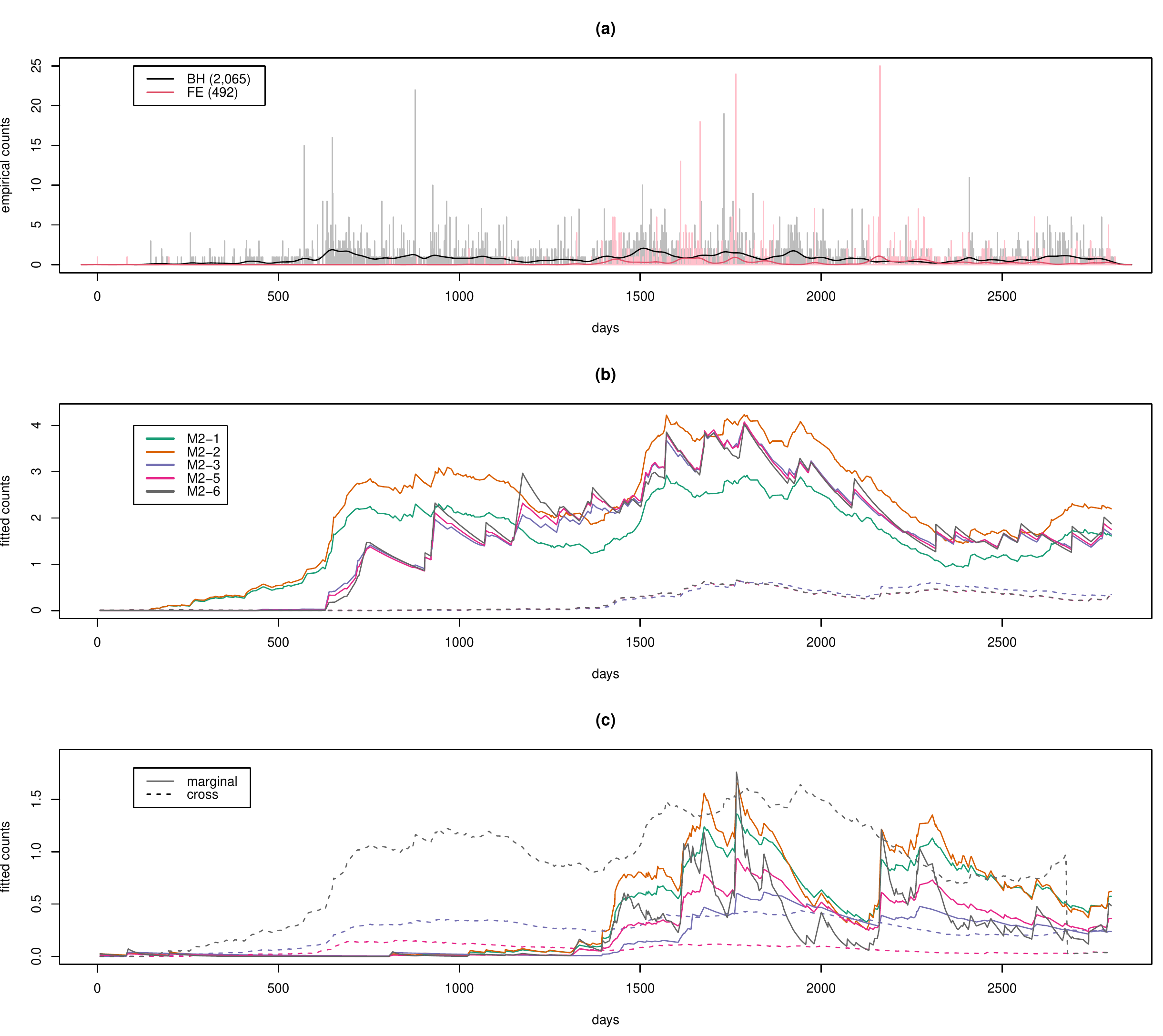}
\caption{Comparison of empirical counts of attacks each day and the corresponding fitted values. (a) Empirical counts of attacks by BH and FE with smoothed curves (b) Fitted values for number of attacks by BH (marginal) and attacks by BH triggered by FE (cross) (c) Similar to (b) except they are for attacks by FE (marginal) and attacks by FE triggered by BH. (b) and (c) share the same legends. In (a), numbers in the legend denote the total number of attacks by each group during the period. See Table~\ref{fitted-numbers} for corresponding fitted numbers from marginal and cross-triggering.}
\label{n-timeseries}
\end{figure}

Considering the performance of these models in regard to the temporal structure, Figure~\ref{n-timeseries} shows comparisons of empirical daily attacks and corresponding fitted values from all models considered. Both empirical counts and fitted {(expected)} values are calibrated such that the y-axis represents the observed (or fitted) counts. Note that there are total 2,065 attacks by BH and 492 by FE in our sample period. Overall, the fitted time series curves for both groups match well with the empirical temporal structure in Figure~\ref{n-timeseries}(a). Note that models with cross-triggering allow us to ``decompose" the triggering structure into marginal and cross-triggering (models M2-3, M2-5, and M2-6). Those models fitted with cross-triggering (over time) differ significantly across different models, especially for FE.


Finally, Table~\ref{fitted-numbers} shows the expected number of attacks attributed to the marginal and cross-triggering structure for each model considered (excluding M2-4, which is omitted due to its unreasonable cross-triggering structure as discussed in Section 4 and the unreasonable estimate of $\alpha_{fb}$ reported in Table 3). Recall that the total number of observed attacks during the sample period was 2,075 for BH and 493 for FE. In Table~\ref{fitted-numbers}, we see that all the models give smaller expected values compared to the actual total number of attacks (marginal plus cross), especially for BH where each of the model's expected value is roughly one-half of the observed counts. This may indicate the need for future development of the base rate structure in subsequent work. In relative terms, however, M2-6 seems to do the best, especially when it comes to predicting the total number of attacks by FE. Interestingly, the ratios between counts due to marginal triggering vs cross-triggering are consistent across different models (M2-3 to M2-6) for Boko Haram, but not for Fulani Extremists. In particular, M2-3 and M2-6 imply that the majority of attacks by FE are due to cross-triggering, while M2-5 does not. Further exploration of how attack patterns may be interrelated is something that should guide future work. 

\begin{table}[h!]
\centering
\caption{Expected number of attacks by marginal and cross-triggering for Nigerian data}
\begin{tabular}{|l||c|c|c|c|c|}
\hline
&M2-1&M2-2&M2-3&M2-5&M2-6\\
\hline
\hline
BH (marginal)&750&1,062&810&823&835\\
BH $\leftarrow$FE (cross)&\cellcolor{gray}&\cellcolor{gray}&119&106&107\\
\hline
Total for BH&750&1,062&929&929&942\\
\hline
\hline
FE (marginal)&201&224&94&126&122\\
FE $\leftarrow$ BH (cross)&\cellcolor{gray}&\cellcolor{gray}&122&32&411\\
\hline
Total for FE&201&224&216&158&533\\

\hline
\end{tabular}
\label{fitted-numbers}
\end{table}

\section{Discussion}
\label{sec:discussion}

In this paper, we develop flexible univariate and bivariate spatio-temporal Hawkes process models suitable for complex point patterns like those resulting from terror attacks. In particular, allowing the spatio-temporal triggering function to depend on covariate data, freeing the cross triggering function to be non-monotonic in spatial lags, and permitting the cross triggering functions to be asymmetric has been proven to be effective in our studies of terrorism in Afghanistan and Nigeria. 

\begin{figure}[hbt!]
\centering
\includegraphics[width=0.45\textwidth]{ 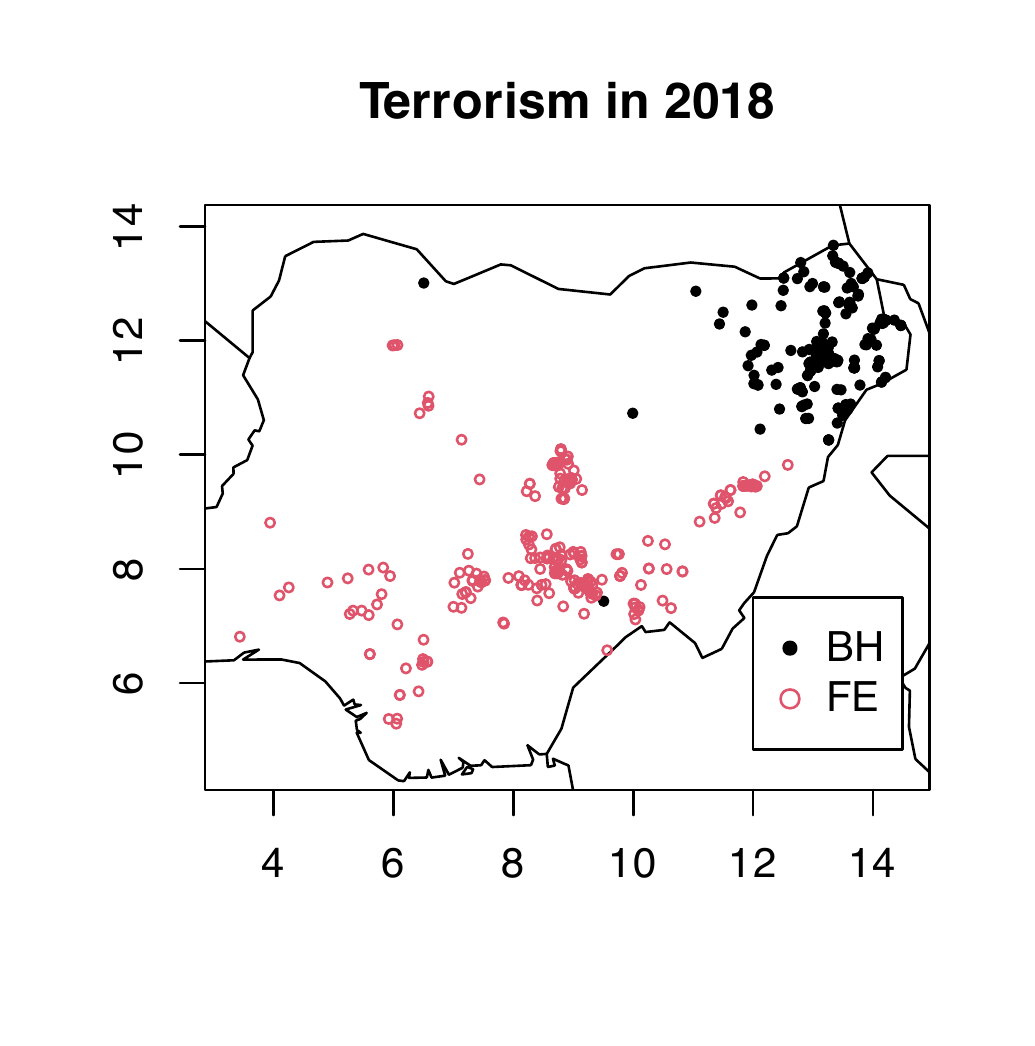}
\caption{{Terrorism attacks by BH and FE in Nigeria during 2018}}
\label{nig-18}
\end{figure}

{We have compared all the models considered based on maximized log-likelihood as well as the three information criteria, the AIC, BIC, and HQ. For the Nigerian data, we believe that one of the main reasons why our proposed models outperform existing models significantly for all three information criteria is due to their flexibility in describing spatial cross-triggering structure. Figure~\ref{nig-18} shows the spatial patterns of attacks by BH and FE in 2018 (we used the data during 2009-2017 to fit the models). Note that a similar spatial separation pattern between the two groups is apparent to what we observed during 2009-2017 period. Furthermore, median (or mean) values of longitudinal lags and latitudinal lags (all in degrees) between location of attacks by BH and FE are 4.41 (or 4.46) and 3.61 (or 3.52), respectively, which essentially match with estimated parameter values for $\eta_c$ and $\xi_c$ in Table~\ref{estimate-new-nigeria-update-new}.}

{In an effort to assess out-of-sample
 performance, we follow 
\cite{ilhan_kozat20} who used negative log-likelihood values of ``test data'' using the fitted results with the ``training data''. We performed similar calculations for Afghanistan data from 2014 (as here our original models were fit to data from 2002-2013) and Nigerian data from 2018 (as here our original models were fit to data from 2009-2017). Note that there were 1,029 attacks by Taliban in Afghanistan during 2014, and 209 and 300 attacks by BH and FE, respectively, in Nigeria during 2018. Table~\ref{test-lik} shows summary results for this analysis, with some models with simpler structure (and poorer performance) excluded in the table for concision. One can see that for both country examples, the best models based on the three information criteria gives the largest log-likelihood values. These are also the best performing models from the training data (i.e., the original analysis samples), indicating that our preferred models did well both in- and out-of-sample based on log-likelihood values.}

\begin{table}[h]
    \centering
     \caption{{Log-likelihood evaluated with the data from 2014 (for Afghanistan) and 2018 (for Nigeria) with parameter values reported in Tables~\ref{afg-estimate} and \ref{estimate-new-nigeria-update-new}. The best log-likelihood values are marked bold-faced.}}
    \begin{tabular}{|c||c|c||c|c|c|c|c|}
    \hline
    &\multicolumn{2}{c||}{Afghanistan 2014}&\multicolumn{5}{c|}{Nigeria 2018}\\
    \hline
  Model  &M1-1&M1-3&M2-2&M2-3&M2-4&M2-5&M2-6\\
    \hline
    \hline
    loglik&4,796.01&{\bf 11,459.29}&833.15&936.81&1,073.67&1,179.29&{\bf 1,193.59}\\
  \hline
    \end{tabular}
         \label{test-lik}
\end{table}

{While we believe that this is a useful first cut for evaluating the out-of-sample performance of our model, we believe that a better approach would include all the information up to the next prediction interval (including subsequent predictions). Further research developing best practices for assessing out-of-sample predictions in Hawkes process models is especially useful for the kinds of social events we consider here (along with cognate research on crime). In future work, other model diagnostics methods to check how well models describe complex spatio-temporal nature of the data as well as their complex joint structure need to be developed (e.g. temporal motifs introduced in \cite{pmlr-v162-soliman22a} for temporal Hawkes processes). Importantly, since in spatio-temporal Hawkes process models the intensity at time $t$ is defined conditionally upon all the events happened up to time $t$, researchers will need to consider how to best use all available information when generating predicted sequences and patterns of events.}

What do our results tell us about patterns of terrorism? First, for univariate analyses we demonstrate the importance of accounting for the spatial distribution of attacks (via spatio-temporal Hawkes process models) rather than aggregating across space as in the temporal Hawkes process models which have predominated in applied statistics research on terrorism. Second, in context with multiple terror groups (e.g., Nigeria), our results demonstrate the clear utility of both: i) accounting for attacks by both (or all) groups, and ii) the importance of \emph{separately} accounting for these groups (i.e., using bivariate rather than univariate models). We show how these issues can be well handled using our generalized version of the spatio-temporal Hawkes Process model, leaving for future work comparisons between this and alternative spatial cluster models. 

There are other complexities and challenges with using point pattern data from the social sciences that we have not addressed here. For example, in Section~\ref{sec:numerical} we briefly mention the problem of potential geolocation errors (or spatial uncertainty) in these data, an issue that requires greater attention in future work. The spatial uncertainty in the locations for these events has been identified elsewhere \citep{weidmann2015accuracy, cookWeidmann2022}, yet there are few solutions and little discussion of the salience of these problems for point pattern analysis. Much of the existing work focuses on particular countries, however, geolocation uncertainty can be particularly problematic for researchers interested in overtime, cross-national analyses. The GTD data we use here, for example, is considered among the most comprehensive (unclassified) database on terrorist events, yet it still suffers from known geospatial inaccuracy. As such, the GTD provides the `specificity' variable \citep{start}, as summarized in Table~\ref{tab:spec}. Figure~\ref{afg-election} earlier showed a spatial map of spatial attack patterns on one day with different specificity values in Afghanistan. 

\begin{table}[hbt!]
    \centering
     \caption{GTD `specificity' description}
    \label{tab:spec}
    \begin{tabular}{|c|l|}
    \hline
        Value &\hspace{4cm} Description  \\
        \hline 
  {1}  & Event occurred in city/village/town and lat/long is for that location\\
         \hline
         \multirow{2}{*}{2}&Event occurred in city/village/town and no lat/long could be found, so coordinates \\
         &are for centroid of smallest subnational administrative region
identified\\
\hline
\multirow{2}{*}{3}&Event did not occur in city/village/town, so coordinates are for centroid of \\
&smallest subnational administrative region identified\\
\hline
\multirow{2}{*}{4}&No 2\textsuperscript{nd} order or smaller region could be identified, so coordinates are for center \\
&of 1\textsuperscript{st} order administrative region\\
\hline
\multirow{2}{*}{5}&No 1\textsuperscript{st} order administrative region could be identified for the location of the attack, \\
&so latitude and longitude are unknown\\
\hline
    \end{tabular}
\end{table}

\begin{figure}[bth!] \includegraphics[angle=0,totalheight=12cm]{ 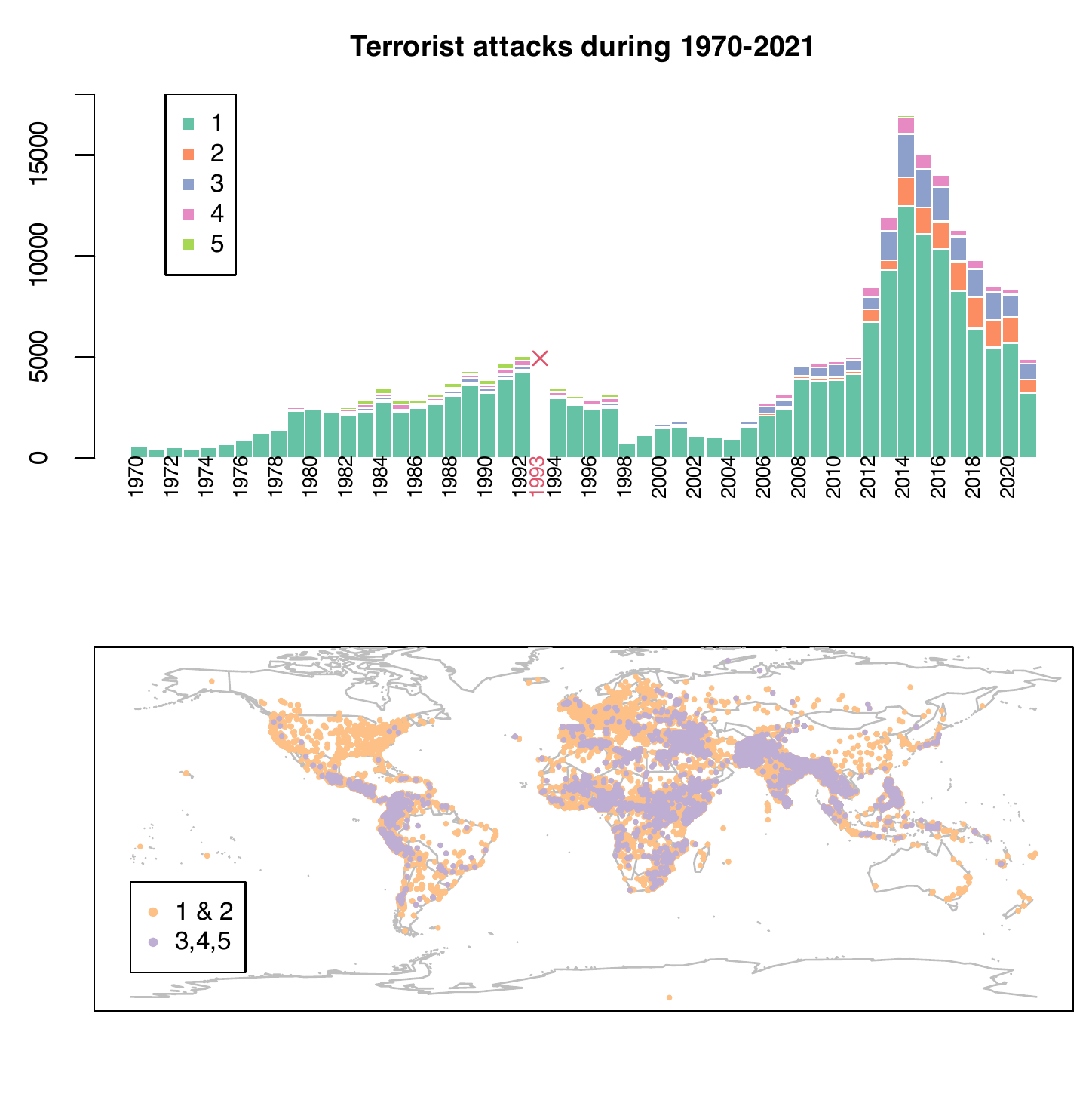}
\caption{{Geocoding specificity (using `specificity') in the GTD. The numbers in the legend indicate the value of specificity. Note that incidents of terrorism from 1993 are not present in the GTD, as they were lost prior to START's compilation of the GTD from multiple sources ({\tt https://www.start.umd.edu/gtd/faq/}). Country level statistics for 1993 are provided in the GTD, and the total number of attacks in 1993 are marked with red x.}}
\label{new-spec}
\end{figure}

Using this information, we generate Figure~\ref{new-spec} to show the annual number of terror attacks for each level of specificity, {as well as spatial locations of these attacks}. {It is evident from the top figure that}, despite (or perhaps because of) the dramatic increase in total attacks globally over the last decade, we have not seen much improvement in geocoding accuracy. Moreover, we clearly see year-to-year variation in the number of attacks imprecisely located (and, to a lesser extent, their proportion to precisely located events). In addition to overtime variation in geospatial accuracy, we also see cross-national differences. {The bottom figure} displays the spatial distribution of all attacks during 1970-2021. {Locations of attacks with more ``certain" locational information, i.e., specificity 1 or 2, are contrasted with those with more ``uncertain" locational information, i.e., specificity 3 and higher.} As expected, developed democracies such as the U.S., Australia, and most of Western Europe report mostly accurate spatial locations (with specificity = 1 or 2), while many countries in the Middle East, Africa, and South America report attacks with less-certain spatial location information. There is also a significant number of attacks with specificity = 5 that are not plotted. This has clear consequences for subsequent analysis, as there would be a clear risk of confounding due to this error if researchers include some determinants that are more/less abundant (ex. GDP, democracy, etc.) in the countries with higher levels of geospatial accuracy.
 
 This data limitation requires more rigorous statistical methods to address the spatial uncertainty issue directly during estimation. Similar problems have been confronted in other research areas \citep[e.g.,][]{zimmerman_et_al07, chakraborty_gelfand10, fanshawe_diggle11, heaton_et_al19}, yet these solutions may not be suitable here. First, many of these models rely on restrictive models of the spatial errors, such as assuming that the two components (that is, $x$ and $y$ coordinates) are uncorrelated. Second, these models often use the same distribution -- typically uniform, normal, or a finite mixture of normal -- with the same parameter values for the two components. Finally, the distributions for the error process are assumed to be common across the entire spatial domain. While these assumptions are reasonable in some settings -- for example, in \cite{heaton_et_al19}'s model of Bronchiolitis incidence the location error was induced by jittering to ensure privacy -- they are not generally satisfied in social science event data, which contain more complex error structure. We believe that location errors are likely to vary across space and/or time as a function of population, economic development, etc. As such, a more general model of the spatial error is necessary for these applications, and we leave this for our future work.

\begin{funding}

The authors acknowledge support by NSF DMS-1925119 and DMS-2123247. Mikyoung Jun also acknowledges support by NIH P42ES027704. 
\end{funding}

\bibliographystyle{imsart-nameyear}
\bibliography{reference-new}

\newpage
\begin{center}
\bf  \Large Appendix
\end{center}

\vspace{3mm}

\noindent {\bf \large Appendix A: Parametrization for stability conditions}\\

For the univariate case, since the triggering functions considered are in the density form, it is straightforward to ensure the stability conditions are met. 

For the bivariate case, the condition discussed in Section~\ref{sec:stability} translates to the condition that the absolute values of eigenvalues of the matrix, 
$\mathrm{A}=\begin{pmatrix} 
\alpha_b & \alpha_{bf}\\
\alpha_{fb} & \alpha_f\\
\end{pmatrix}
,$ do not exceed 1. We also notice from the BH and FE data that BH has a lot more events (which may result in stronger marginal as well as cross-triggering). 
To achieve this, we first start with a matrix $$\tilde{A}=\begin{pmatrix} \cos{\theta} & -\sin{\theta} \\ \sin{\theta} & \cos{\theta}
\end{pmatrix} \begin{pmatrix} \lambda_b & 0 \\ 0 &\lambda_f
\end{pmatrix}{\begin{pmatrix} \cos{\theta} & -\sin{\theta} \\ \sin{\theta} & \cos{\theta}
\end{pmatrix} }^{-1}$$
with $-\pi/2 \leq \theta\leq\pi/2 $, and $0\leq\lambda_f\leq \lambda_b\leq 1$. Since $\tilde{A}$ is symmetric, but we allow $\alpha_{bf} \neq \alpha_{fb}$, we set $\alpha_b=\tilde{A}_{1,1} $, $\alpha_f=\tilde{A}_{2,2}$, $\alpha_{bf} = \tilde{A}_{1,2}+b$ and $\alpha_{fb}=\tilde{A}_{2,1}$ for a small $b$: $0 \leq b \leq 1-\lambda_b$. A small constant $b$ is introduced in order to allow asymmetry of $\mathrm{A}$, that is, $\alpha_{bf} \geq  \alpha_{fb}$. The four parameters, $\theta$, $\lambda_b$, $\lambda_f$, and $b$, are estimated along with the rest of parameters in the model. Exponential, and logit transformations are used for these parameters within the objective function to ensure that these parameter values stay within the defined domain. Also, every iteration for numerical optimization, eigenvalues of $\mathrm{A}$ have been checked, and we made sure that they satisfy the stability conditions.

\vspace{5mm}

\noindent {\bf \large Appendix B: Numerical optimizations and calculation of asymptotic standard errors}\\

Numerical optimization of log-likelihood function was done using {\it optim} and {\it nlm} functions in {\sf R} extensively. For each model, both procedures were used multiple times to ensure that we reach the numerical optimum. Furthermore, Hessian matrices were checked to ensure proper convergence of the iteration. 

Hessian matrices were also utilized to achieve asymptotic standard errors of parameter estimates. However, as described in Appendix A, many of the parameters were transformed in the log-likelihood functions to ensure that parameters stay in the right range (for instance, the triggering distance parameters were transformed with an exponential function to ensure positivity). With those parameters transformed within log-likelihood function, proper Jacobian needs to be applied to Hessian matrices to obtain correct standard errors. However, the transformation performed for parameters $\alpha_b$, $\alpha_f$, $\alpha_{bf}$, and $\alpha_{fb}$ are rather complex. Therefore, for the bivariate models with these four parameters, we used {\it numDeriv} package in {\sf R} to calculate Hessian matrices (at the converged parameter values) directly to avoid the need to apply Jacobians.

\end{document}